\documentclass[aps,prd,12pt,notitlepage,nofootinbib,tightenlines]{revtex4-1}
\usepackage{amsmath}
\usepackage{amssymb}
\usepackage{epsfig}
\usepackage{graphicx}
\usepackage{bm}
\usepackage{times}
\usepackage{braket}
\usepackage{color}
\usepackage{slashed}
\usepackage{hyperref}
\definecolor{Red}{rgb}{1.,0.,0.}

\definecolor{Blue}{rgb}{0.,0.,1.}

\definecolor{nicered}{rgb}{0.7,0.1,0.1}
\definecolor{nicegreen}{rgb}{0.1,0.5,0.1}

\hypersetup{colorlinks,citecolor=nicegreen,linkcolor=nicered}
\begin{document}
\newcommand{\beq}{\begin{eqnarray}}
\newcommand{\eeq}{\end{eqnarray}}
\newcommand{\non}{\nonumber\\ }

\newcommand{\jpsi}{J/\Psi}
\newcommand{\ppa}{\phi_\pi^{\rm A}}
\newcommand{\ppp}{\phi_\pi^{\rm P}}
\newcommand{\ppt}{\phi_\pi^{\rm T}}

\newcommand{\zerot}{ {\textbf 0_{\rm T}} }
\newcommand{\tr} {{\rm Tr} }
\newcommand{\kt}{k_{\rm T} }
\newcommand{\fb}{f_{\rm B} }
\newcommand{\fk}{f_{\rm K} }
\newcommand{\mb}{m_{\rm B} }
\newcommand{\mw}{m_{\rm W} }
\newcommand{\im}{{\rm Im} }

\newcommand{\acp}{{\cal A}_{\rm CP}}
\newcommand{\pb}{\phi_{\rm B}}

\newcommand{\xeba}{\bar{x}_2}
\newcommand{\xsba}{\bar{x}_3}
\newcommand{\peas}{\phi^A}

\newcommand{\pvsl}{ p \hspace{-2.0truemm}/_{K^*} }
\newcommand{\esl}{ \epsilon \hspace{-2.1truemm}/ }
\newcommand{\psl}{ p \hspace{-2truemm}/ }
\newcommand{\ksl}{ k \hspace{-2.2truemm}/ }
\newcommand{\lsl}{ l \hspace{-2.2truemm}/ }
\newcommand{\nsl}{ n \hspace{-2.2truemm}/ }
\newcommand{\vsl}{ v \hspace{-2.2truemm}/ }
\newcommand{\epsl}{\epsilon \hspace{-1.8truemm}/\,  }
\newcommand{\bfkk}{{\bf k} }
\newcommand{\calm}{ {\cal M} }
\newcommand{\calh}{ {\cal H} }
\newcommand{\calo}{{\cal O}}

\def \appb{{\bf Acta. Phys. Polon. B}}
\def \cpc{ {\bf Chin. Phys. C}}
\def \ctp{ {\bf Commun. Theor. Phys. }}
\def \epjc{{\bf Eur. Phys. J. C}}
\def \jhep{{\bf J. High Energy Phys. }}
\def \jpg{ {\bf J. Phys. G} }
\def \mpla{{\bf Mod. Phys. Lett. A}}
\def \npb{ {\bf Nucl. Phys. B}}
\def \plb{ {\bf Phys. Lett. B}}
\def \ppnp{{\bf Prog. part. $\&$ Nucl.Phys.} }
\def \pr{  {\bf Phys. Rep.} }
\def \prc{ {\bf Phys. Rev. C}}
\def \prd{ {\bf Phys. Rev. D}}
\def \prl{ {\bf Phys. Rev. Lett.}  }
\def \ptp{ {\bf Prog. Theor. Phys. }}
\def \zpc{ {\bf Z. Phys. C}}


\title{The NLO twist-3 contributions to $B \to \pi$ form factors
in $k_{T}$ factorization}
\author{Shan  Cheng$^{1}$}
\author{Ying-Ying Fan$^{1}$}
\author{Xin Yu$^2$}
\author{Cai-Dian L\"u$^2$}\email{lucd@ihep.ac.cn}
\author{Zhen-Jun Xiao$^{1,3}$}\email{xiaozhenjun@njnu.edu.cn}
\affiliation{1.  Department of Physics and Institute of Theoretical Physics,
Nanjing Normal University, Nanjing, Jiangsu 210023, People's Republic of China,}
\affiliation{2.  Institute of High Energy Physics and Theoretical
Physics Center for Science  Facilities, Chinese Academy of Sciences,
Beijing 100049, People's Republic of China,}
\affiliation{3. Jiangsu Key Laboratory for Numerical Simulation of Large Scale Complex Systems,
Nanjing Normal University, Nanjing 210023, People's Republic of China}
\date{\today}
\begin{abstract}
In this paper, we calculate the next-to-leading-order (NLO) twist-3 contribution
to the form factors of $B \to \pi$ transitions by employing
the $k_{T}$ factorization theorem.
All the infrared divergences regulated by the logarithms $\ln(k_{iT}^{2})$ cancel
between those from the quark diagrams and from the effective diagrams for the initial
$B$ meson wave function and the final pion meson wave function.
An infrared finite NLO hard kernel is therefore obtained, which confirms
the application of the $k_{T}$ factorization theorem to $B$ meson semileptonic decays
at twist-3 level.
From our analytical and numerical evaluations, we find that
the NLO twist-3 contributions to the form factors $f^{+,0}(q^2)$ of $B \to \pi$ transition
are similar in size, but have an opposite sign with the NLO twist-2 contribution,
which leads to a large cancelation between these two NLO parts.
For the case of $f^+(0)$, for example,
the $24\%$ NLO twist-2 enhancement to the full LO prediction
is  largely canceled by the negative ( about $-17\%$ ) NLO twist-3 contribution,
leaving a small and stable $7\%$ enhancement to the full LO prediction in
the whole range of $0\leq q^2\leq 12$ GeV$^2$.
At the full NLO level, the perturbative QCD prediction is $F^{B \to \pi}(0)=0.269^{+0.054}_{-0.050}$.
We also studied the possible effects on the pQCD predictions when different sets of the B meson and
pion distribution amplitudes are used in the numerical evaluation.
\end{abstract}



\maketitle

\section{Introduction}

Without end-point singularity, $k_{T}$ factorization theorem
\cite{npb325-62,npb360-3,prl74-4388} is a better tool to deal
with the small $x$ physics when comparing with other factorization
approaches\cite{plb87-359,prd22-2157,plb94-245,qcd1993,npb685-249}.
Based on the $k_{T}$ factorization theorem, perturbative QCD (pQCD)
factorization approach\cite{plb504-6,prd63-074009,prl65-2343,li2003} is
a successful factorization approach to handle the heavy to light exclusive
decay processes.
As an effective factorization theorem, the $k_{T}$ factorization
should be valid at every order expanded by strong coupling $O(\alpha^{n}_{s})$,
where $n$ is the power of the expansion.

Recently, the next-to-leading-order(NLO) twist-2 (the leading twist) contributions
to the form factors for the $\pi \gamma^{\star} \to \gamma$, $\pi \gamma^{\star}
\to \pi$ and $B \to \pi$ transitions have been evaluated
\cite{prd76-034008,prd83-054029,prd85-074004}
by employing the  $\kt$ factorization theorem \cite{npb325-62,npb360-3,prl74-4388},
and an infrared finite $k_T$ dependent hard kernel were obtained at the NLO level
for each considered process. It is worth of mentioning that a new progress
about pion form factor in the $\pi \gamma^* \to \gamma$ scattering
has been made in Ref.~\cite{jhep1401-004} very recently, where the authors made
a joint resummation for the pion wave function and the pion transition form factor
and proved that the $\kt$ factorization is scheme independent.
These NLO contributions could produce sizable effects to the LO hard kernels.
For example, the NLO twist-2 contribution to the form factor $F_0^{B\to \pi}(0)$
for $B \to \pi$ transition can provide $\sim 30\%$ enhancement to the corresponding
full LO form factor \cite{prd85-074004}.
In a recent paper \cite{cheng14a}, we calculated the NLO twist-3 contribution
to the pion electromagnetic form factor $F_{\pi\gamma}(Q^2)$  in the
$\pi \gamma^{\star} \to \pi$ process
by employing the $\kt$ factorization theorem, and found infrared finite
NLO twist-3 corrections to the full LO hard kernels \cite{cheng14a}.

In this paper, following the same procedure of Ref.~\cite{prd85-074004},
we will calculate the NLO twist-3 contribution to the form factor
of $B \to \pi$ transition, which is the only missing piece at the NLO level.
The light partons are also considered to be off-shell by $\kt^2$ in both
QCD quark diagrams and effective diagrams for hadron wave functions.
The radiation gluon from the massive $b$ quark generates the soft divergence only.
Such soft divergence can be regulated either by the virtuality of internal
particles or by the virtuality $\kt^2$ of other light partons, to
which the emission gluons were attached.
So we can replace the off-shell scale $\kt^{2}$ for the light parton
by $m_{g}$ for the massive b quark safely to regulate the IR divergences
from the massive $b$ quark line, where $m_{g}$ means the mass of the
gluon radiated from the b quark. That means, the b quark remains on-shell in the framework.

We will prove that the IR divergences in the NLO QCD quark
diagrams could be canceled by those in the effective diagrams, i.e., the
convolution of the $O(\alpha_{s})$ $B$ meson and $\pi$ meson wave functions
with the LO hard kernel. The IR finiteness and $\kt$-dependent NLO hard kernel
were also derived at the twist-3 level for the $B\to\pi$ transition form
factor, which confirms the application of the $\kt$ factorization
theorem to $B$ meson semileptonic decays at both the twist-2 and twist-3 level.

In our calculation for the NLO twist-3 contribution, the resummation
technology\cite{prd66-094010,plb555-197} is applied to
deal with the large double logarithms $\alpha_{s}\ln^2\kt$ and
$\alpha_{s}\ln^{2}x_i$, where $x_{i}$ being the parton momentum
fraction of the anti-quark in the meson wave functions.
With appropriate choices of $\mu$ and $\mu_{f}$, say being lower than
the $B$ meson mass, the NLO corrections are under control.
From numerical evaluations we find that the NLO correction at twist-3
is about $-17\%$ of the LO part, while the NLO twist-2 contribution can provide
a $24\%$ enhancement to the LO one. This means that the NLO twist-2 contribution
to the form factor $F^{B\to \pi}(0)$ are largely canceled by the NLO twist-3 one, leaves a net small
correction to the full LO form factor, around or less than $7\%$ enhancement.

The paper is organized as follows. In Sec.~II, we give a brief introduction for the
calculations of the LO diagrams relevant with the form factor of $B\to\pi$ transition.
In Sec.~III, we calculate the NLO twist-3 contribution to the $B \to \pi$ form factor.
The relevant $O(\alpha^{2}_{s})$ QCD quark diagrams are calculated
analytically, the convolutions of $O(\alpha_{s})$ wave
functions and $O(\alpha_{s})$ hard kernel are made in the same way
as those for the evaluation of the NLO twist-2 contribution.
And finally we extract out the expression of the factor
$F_{\rm NLO-T3}^{B\to \pi}(x_i,\mu,\mu_f,\eta)$, which describes the NLO twist-3
contribution to the form factor $F^{B\to \pi}(x_i,\mu,\mu_f,\eta)$.
In Sec.~IV we calculate and present the numerical results for
the relevant form factors and examine the $q^2$-dependence of $F^+(q^2)$
and $ F^0(q^2)$ at the LO and NLO level, respectively. A short summary
was given in the final section.


\section{LO analysis}

By employing the $\kt$ factorization theorem, the LO twist-2 and twist-3 contributions
to the  form factor of $B\to \pi$ transition have been calculated many years
ago \cite{plb504-6,prd63-074009,prl65-2343,li2003}. For the sake of the readers, we here
present the expressions of the leading order hard kernels directly.

\begin{figure}[tb]
\vspace{-3cm}
\begin{center}
\leftline{\epsfxsize=14cm\epsffile{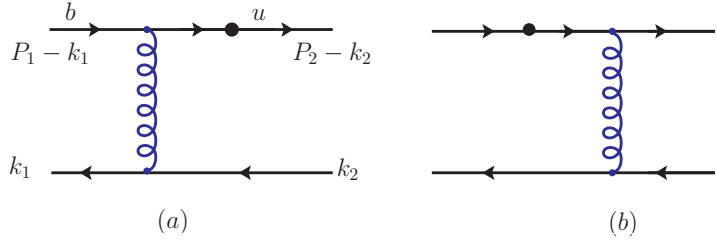}}
\end{center}
\vspace{-15cm}
\caption{Leading-order quark diagrams for the $B\to\pi$ transition form
factor with symbol $\bullet$ representing the weak vertex of $B \to \pi l\bar{\nu}_l$
decay.}
\label{fig:fig1}
\end{figure}

The $B\to\pi$ transition form factors are defined via the matrix element
\beq
<\pi(p_2)|\bar{u}\gamma^{\mu}b|B(p_1)> &=&
f^{+}(q^{2})(p^{\mu}_{1}+p^{\mu}_{2})
+ [f^{0}(q^{2})-f^{+}(q^{2})]\frac{m^{2}_{B}-m^{2}_{\pi}}{q^{2}}q^{\mu} ,
\label{eq:ffme}
\eeq
where $m_{B}~(m_{\pi})$ is the $B~(\pi)$ meson mass, and $q = p_1 - p_2$
is the transfer momentum.
The momentum $p_1~(p_2)$ is chosen as $p_1 = p^{+}_{1}(1,1,{\textbf
0_{T}})~(p_2 = (0,p^{-}_{2},{\textbf 0_{T}}))$ with the component
$p^{+}_{1} = m_{B}/\sqrt{2}$ and $p^{-}_{2} = \eta m_{B}/\sqrt{2}$.
Here the parameter $\eta=1-q^2/m_B^2$ represents the energy fraction
carried by the pion meson, and $\eta \sim O(1)$ when in the large
recoil region of pion.
According to the $\kt$ factorization, the anti-quark $\overline{q}$
carries momentum $k_{1} = (x_{1}p^{+}_{1}, 0 , \bfkk_{\rm 1T})$ in the $B$
meson and $k_{2} = (0, x_{2}p^{-}_{2}, \bfkk_{\rm 2T})$ in the pion meson
as labeled in Fig.~\ref{fig:fig1}, $x_{1}$ and $x_{2}$ being the
momentum fractions.
The follow hierarchy is postulated in the small-x region:
\beq
m^{2}_{B} \gg x_2 m^2_B \gg x_1 m^2_B \gg x_1 x_2 m^2_B , k^2_{1T}, k^2_{2T},
\label{eq:hierarchy}
\eeq
which is roughly consistent with the order of magnitude: $x_1\sim 0.1$, $x_2\sim 0.3$,
$m_B\sim 5$ GeV, and $\kt \lesssim 1$ GeV \cite{prd85-074004}.

The LO hard kernels are obtained after sandwiching Fig.~1 with the $B$ meson
and the pion meson wave functions\cite{plb504-6,prd63-074009,npb592-3}
\beq
&&\Phi_{B}(x_1,p_1)=\frac{1}{2\sqrt{N_{c}}}
\left (\psl_{1} +m_{B} \right ) \gamma_5 \left [ \nsl_{+}\phi^{+}_B(x_{1})
+\left (\nsl_{-}-k^{+}_{1}\gamma^{\nu}_{\bot}\frac{\partial}{\partial
\textbf{k}^{\nu}_{1T}} \right )\phi^{-}_B(x_{1}) \right ],\label{eq:phiB}\\
&&\Phi^{\rm T2}_{\pi}(x_2,p_2)=\frac{1}{\sqrt{2N_{c}}}\gamma_5
\psl_{2} \phi_{\pi}^A(x_{2}),\label{eq:phipit2}\\
&&\Phi^{\rm T3}_{\pi}(x_2,p_2)=\frac{1}{\sqrt{2N_{c}}}
m_{0}\gamma_{5}\left [\phi^{P}_{\pi}(x_{2})
-(\nsl_{-}\nsl_{+}-1)\phi^{T}_{\pi}(x_{2}) \right ],
\label{eq:phipit3}
\eeq
where $m_0$ is the chiral mass of pion,
$\Phi^{\rm T2}_{\pi}$ and $\Phi^{\rm T3}_{\pi}$ denote the pion meson wave function
at twist-2 and twist-3 level, the dimensionless  vectors are defined by
$\nsl_{+} = (1,0,\textbf{0}_T)$, and $\nsl_{-} = (0,1,\textbf{0}_T)$,
and $N_{c}$ is the number of colors.
Without considering the transverse component of the $B$ meson spin projector,
the LO twist-3 contribution for Fig.~1(a) is of the form,
\beq
&&H^{(0)}_{\rm a, T3}(x_{1},k_{1T},x_{2},k_{2T}) = \frac{g^{2}_{s} C_{F} \; m_{0} m_{B}}
     {[(p_1-k_{2})^{2}-m^{2}_{B}][(k_{1}-k_{2})^{2}]} \non
&&~~~~~~\cdot \left \{ \phi^{P}_{\pi}(x_{2}) \left [\phi^{+}_{B}(x_{1}) (4 \frac{p_2^{\mu}}{\eta} - 4 x_2 p_2^{\mu})
                                +\phi^{-}_{B}(x_{1}) (4 p_1^{\mu} - 4 x_2 p_2^{\mu} - 4 \frac{p_2^{\mu}}{\eta}) \right]\right. \non
&&~~~~~~\left.      + \phi^{T}_{\pi}(x_{2}) \left [\phi^{+}_{B}(x_{1}) (4 \frac{p_2^{\mu}}{\eta} - 4 x_2 p_2^{\mu})
                                +\phi^{-}_{B}(x_{1}) (4 \frac{p_2^{\mu}}{\eta} - 4 p_1^{\mu} - 4 x_2 p_2^{\mu})\right ] \right \},
\label{eq:lot3hka}
\eeq
and for Fig.~1(b) we find
\beq
H^{(0)}_{\rm b, T3}(x_{1},k_{1T},x_{2},k_{2T}) = \frac{2g^{2}_{s} C_F\; m_{0}
m_{B}\phi^{P}_{\pi}(x_{2})}{[(p_2-k_{1})^{2}][(k_{1}-k_{2})^{2}]}
\left [ 4p^{\mu}_{2} \phi^{+}_{B}(x_{1}) - 4x_{1}p^{\mu}_{1} \phi^{-}_{B}(x_{1}) \right ],
\label{eq:lot3hkb}
\eeq
where $C_F=4/3$ is the color factor.

The LO twist-2 contributions for Fig.~1(a) and 1(b) are of the form,
\beq
&&H^{(0)}_{\rm a,T2}(x_{1},k_{1T},x_{2},k_{2T}) =  -4g^{2}_{s} C_F m^{2}_{B}
\phi^{A}_{\pi}(x_{2}) \frac{p^{\mu}_{2}  \phi^{-}_{B}(x_{1}) + k^{\mu}_{2}
 \phi^{+}_{B}(x_{1})}{[(p_1-k_{2})^{2}-m^{2}_{B}][(k_{1}-k_{2})^{2}]},
\label{eq:lot2hka}\\
&&H^{(0)}_{\rm b, T2}(x_{1},k_{1T},x_{2},k_{2T}) =  -4g^{2}_{s} C_F m^{2}_{B}
\phi^{A}_{\pi}(x_{2}) x_{1} \frac{(\eta p^{\mu}_{1} - p^{\mu}_{2}) \phi^{+}_{B}(x_{1})
+ p^{\mu}_{2} \phi^{-}_{B}(x_{1})}{[(p_2-k_{1})^{2}][(k_{1}-k_{2})^{2}]}.
\label{eq:lot2hkb}
\eeq

For the LO twist-2 hard kernel $H^{(0)}_{\rm b, T2}$,
it is strongly suppressed by the small $x_2$, as can be seen easily from
Eqs.~(\ref{eq:lot2hka},\ref{eq:lot2hkb}),
and therefore the $H^{(0)}_{\rm a, T2}$ from Fig.~1(a) is the dominant part of the full
LO twist-2 contribution.
Consequently, it is reasonable to consider the NLO twist-2 contributions from
Fig.~\ref{fig:fig1}(a) only in the calculation for the NLO twist-2 contributions.

For the LO twist-3 hard kernel $H^{(0)}_{\rm b, T3}$,  the first term
proportional to $p_2^\mu \phi_B^+(x_1)$ in Eq.~(\ref{eq:lot3hkb}) provides
the dominant contribution, while the second term proportional to
$x_1 p_1^\mu \phi_B^-0(x_1)$ is strongly suppressed by the small $x_1$.
The $H^{(0)}_{\rm a, T3}$ can be neglected safely when compared with $H^{(0)}_{\rm b, T3}$,
due to the strong suppression of small $x_{1}$. We therefore consider only the
$\phi^{P}_{\pi}(x_{2})$ component  in Eq.~(\ref{eq:lot3hkb}) from Fig.~\ref{fig:fig1}(b)
in our estimation for the NLO twist-3 contribution.

The LO hard kernels as given in Eqs.(\ref{eq:lot3hka}-\ref{eq:lot2hkb}) are consistent with
those as given in Refs.~\cite{prd65-014007,epjc23-275}, where the $B$ meson wave
function was defined as
\beq
- \frac{1}{\sqrt{2 N_{c}}}(\psl_{1} +m_{B}) \gamma_5 \left [\phi_{B}(x_{1})
- \frac{\nsl_{+} - \nsl_{-}}{\sqrt{2}} \overline{\phi}_{B}(x_{1}) \right ],
\label{eq:wfB}
\eeq
with the relations
\beq
\phi_{B}=\frac{1}{2} (\phi^{+}_{B} + \phi^{-}_{B}), ~~~~~
\overline{\phi}_{B}=\frac{1}{2} (\phi^{+}_{B} - \phi^{-}_{B}).
\label{eq:wfrela}
\eeq

By comparing the hard kernel $H^{(0)}_{\rm b,T3}$ in Eq.~(\ref{eq:lot3hkb})
with $H^{(0)}_{\rm a,T2}$ in Eq.~(\ref{eq:lot2hka}), one can find that
the LO twist-3 contribution is enhanced by the factor $1/x_1$
and the pion chiral mass $m_0^\pi > 1$, and consequently larger than the
LO twist-2 contribution which are associated with the factor $1/x_2$.
The numerical results of Eqs.~(\ref{eq:lot3hkb},\ref{eq:lot2hka}) in the
large recoil region also show that the LO twist-3 contribution
is larger than the LO twist-2 part, by a ratio of around $60\%$ over  $40\%$.
This fact means that the NLO twist-3 contribution  may be important
when compared with the corresponding NLO twist-2 one, this is one of the
motivations for us to make the evaluation for
the NLO twist-3 contribution to the $B \to \pi$ form factor.

\section{NLO corrections}

Since the dominant NLO twist-3 contribution to the form factor of $B\to \pi$
transition is proportional to the $\phi^{P}_{\pi}(x_2) \phi^{+}_{B}(x_1)$
from the Fig.~\ref{fig:fig1}(b), we here consider only the NLO corrections to
the Fig.~\ref{fig:fig1}(b) coming from the quark-level corrections
and the wave function corrections at twist-3 level,
to find the NLO twist-3 contribution to the form factor of $B\to \pi$ transition.

Under the hierarchy in Eq.~(\ref{eq:hierarchy}), only terms that don't
vanish in the limits of $x_i \to 0$ and $k^2_{iT} \to 0$ are kept to simplify
the expressions of the NLO twist-3 contributions greatly.

\subsection{NLO Corrections from the QCD Quark Diagrams}

\begin{figure}[tb]
\vspace{-2cm}
\begin{center}
\leftline{\epsfxsize=11cm\epsffile{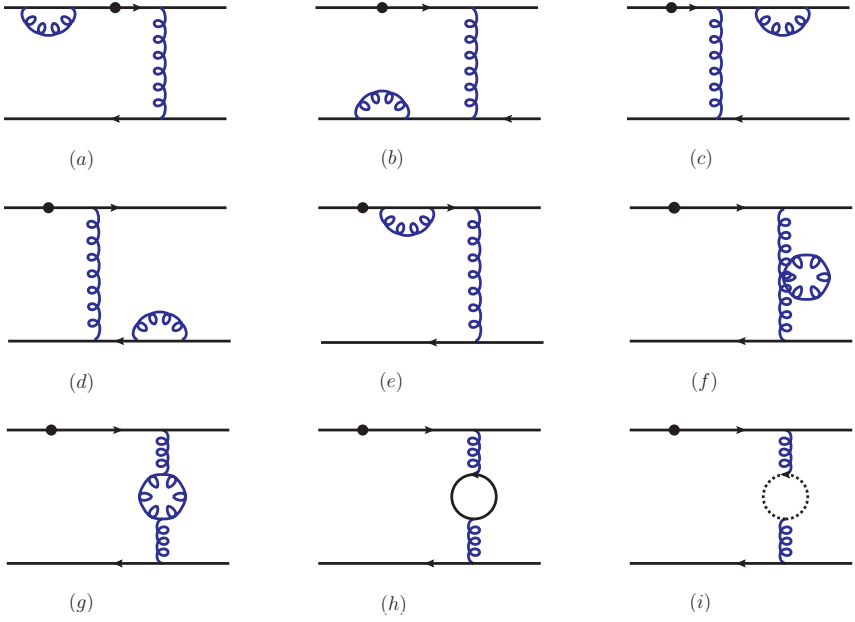}}
\end{center}
\vspace{-7cm}
\caption{Self-energy corrections to fig.1(b).}
\label{fig:fig2}
\end{figure}

\begin{figure}[tb]
\vspace{-2cm}
\begin{center}
\leftline{\epsfxsize=11cm\epsffile{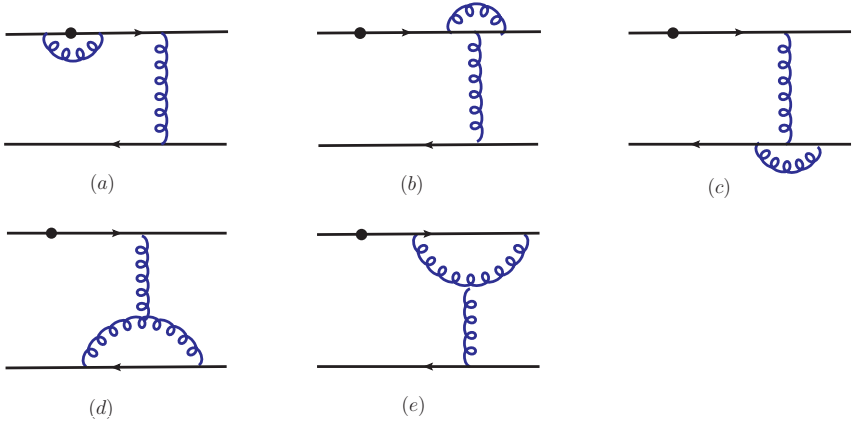}}
\end{center}
\vspace{-9.5cm}
\caption{Vertex corrections to fig.1(b).}
\label{fig:fig3}
\end{figure}

\begin{figure}[tb]
\vspace{-2cm}
\begin{center}
\leftline{\epsfxsize=12cm\epsffile{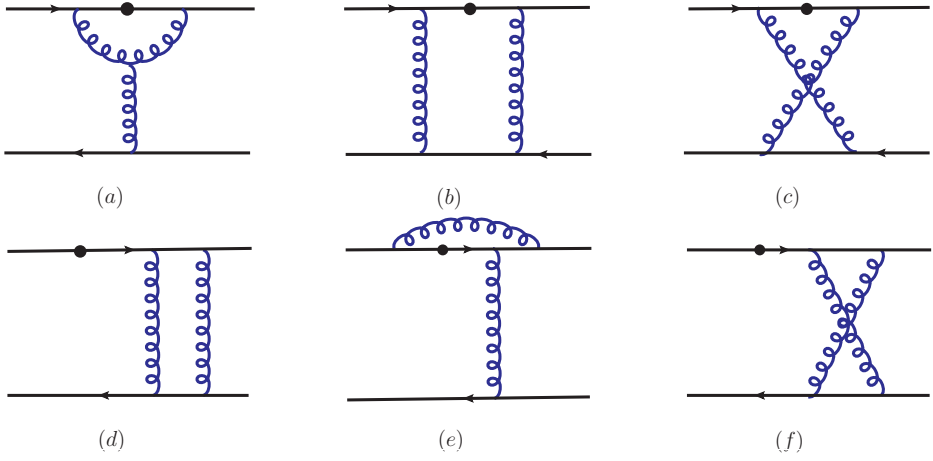}}
\end{center}
\vspace{-10cm}
\caption{Box and pentagon corrections to fig.1(b).}
\label{fig:fig4}
\end{figure}

The NLO corrections to Fig.~\ref{fig:fig1}(b) at quark-level contain
the self-energy diagrams, the vertex diagrams and the box and pentagon
diagrams, as illustrated by  Fig.~\ref{fig:fig2},\ref{fig:fig3},
\ref{fig:fig4}, respectively.
The ultraviolet(UV) divergences are extracted in the dimensional
reduction\cite{plb84-193} in order to avoid the ambiguity from handling
the matrix $\gamma_{5}$. The infrared(IR) divergences are identified
as the logarithms $\ln{m_g}$, $\ln{\delta_1}$ , $\ln{\delta_2}$ and
their combinations, where the dimensionless ratios are adopted,
\beq
\delta_1 = \frac{k^2_{1T}}{m^2_B}, ~~~ \delta_2 = \frac{k^2_{2T}}{m^2_B},
\delta_{12} = \frac{-(k_1 - k_2)^2}{m^2_B},
\label{eq:defi1}
\eeq

By analytical evaluations for the Feynman diagrams as shown in Fig.~\ref{fig:fig2},
we find the self-energy corrections from the nine diagrams:
\beq
&&G^{(1)}_{2a} = -\frac{\alpha_s C_f}{4 \pi}\left [\frac{6}{\delta_1}
\left(\frac{1}{\epsilon} + \ln{\frac{4 \pi \mu^2}{m^2_B e^{\gamma_E}}}
+ \frac{1}{3} \right)
+ \frac{1}{2} \left (\frac{1}{\epsilon} + \ln{\frac{4 \pi \mu^2}{m^2_B e^{\gamma_E}}}
+ 2 \ln{\frac{m^2_g}{m^2_B}} +2 \right) \right ]H^{(0)}, \label{eq:g2a}\\
&&G^{(1)}_{2b} = -\frac{\alpha_s C_f}{8 \pi}\left [\frac{1}{\epsilon} +
\ln{\frac{4 \pi \mu^2}{\delta_1 m^2_B e^{\gamma_E}}} + 2 \right ]H^{(0)},\label{eq:g2b}\\
&&G^{(1)}_{2c,2d} = -\frac{\alpha_s C_f}{8 \pi}\left [\frac{1}{\epsilon} +
\ln{\frac{4 \pi \mu^2}{\delta_2 m^2_B e^{\gamma_E}}} + 2\right ]H^{(0)},\label{eq:g2cd}\\
&&G^{(1)}_{2e} = -\frac{\alpha_s C_f}{4 \pi}\left [\frac{1}{\epsilon} +
\ln{\frac{4 \pi \mu^2}{x_1 \eta  m^2_B e^{\gamma_E}}} + 2\right ]H^{(0)},
\label{eq:g2e}\\
&&G^{(1)}_{2f+2g+2h+2i} = \frac{\alpha_s}{4 \pi}
\left [ \left (\frac{5}{3} N_c -
\frac{2}{3} N_f \right ) \left (\frac{1}{\epsilon} + \ln{\frac{4 \pi \mu^2}{\delta_{12}
m^2_B e^{\gamma_E}}} \right ) \right ]H^{(0)},
\label{eq:sumf2}
\eeq
where $1 / \epsilon$ represents the UV pole, $\mu$ is the renormalization
scale, $\gamma_E$ is the Euler constant, $N_c$ is the number of quark color,
$N_f$ is the number of the quarks flavors, and $H^{(0)}$ denotes the
first term of the LO twist-3 contribution
$H^{(0)}_{\rm b,T3}(x_{1},k_{1T},x_{2},k_{2T})$ as given in Eq.~(\ref{eq:lot3hkb}),
\beq
H^{(0)}(x_{1},k_{1T},x_{2},k_{2T}) = -4g^{2}_{s} C_F\;
m_{0} m_{B} \phi^{P}_{\pi}(x_{2})
\frac{2p^{\mu}_{2} \phi^{+}_{B}(x_{1})}{(p_2-k_{1})^2 (k_{1}-k_{2})^2}.
\label{eq:lohk}
\eeq
It's easy to see that, besides $G^{(1)}_{2e}$ for the subdiagram
Fig.~\ref{fig:fig2}(e), the NLO self-energy corrections listed in
Eqs.~(\ref{eq:g2b},\ref{eq:g2cd},\ref{eq:sumf2}) are identical to the
self-energy corrections for the NLO twist-2 case as given in Eqs.~(7-8,11)
in Ref.~\cite{prd85-074004}.
Except for a small difference in constant numbers, the $G^{(1)}_{2a}$ for the
subdiagram Fig.~\ref{fig:fig2}(a) in Eq.~(\ref{eq:g2a})
is the same one as that as given in Eq.~(7) of Ref.~\cite{prd85-074004} for the
case of the NLO twist-2 contributions.
The reason for such high similarity is that the self-energy diagrams
don't involve the loop momentum flowed into the hard kernel.
Only the Fig.~\ref{fig:fig2}(a), the self-energy correction of the b
quark is emphasized here.
The first term in the square brackets of $G^{(1)}_{2a}$ required the
mass renomalization, and the finite piece of the first term is then
absorbed into the redefinition the b quark mass, with the relation
$(p_1 - k_1)^2 - m^2_B = -k^2_{1T}$.
The second term in the square brackets of $G^{(1)}_{2a}$ represents
the correction to the b quark wave function.
The involved soft divergence is regularized by the gluon mass $m_g$
because the valence b quark is considered on-shell, and the additional
regulator $m_g$ will be canceled by the corresponding soft divergence
in the effective diagrams Fig.~\ref{fig:fig5}(a).
Comparing with the NLO twist-2 case, the result from the subdiagram
Fig.~\ref{fig:fig2}(e) at twist-3 is simple, since it's
the self-energy correction to the massless internal quark line in the
twist-3 case.

By analytical evaluations for the Feynman diagrams as shown in Fig.~\ref{fig:fig3},
we find the vertex corrections from the five vertex diagrams:
\beq
&&G^{(1)}_{3a} =  \frac{\alpha_s C_f}{4 \pi} \left [\frac{1}{\epsilon}
+ \ln{\frac{4 \pi \mu^2}{m^2_B e^{\gamma_E}}} -
\ln^2{x_1} - 2 \ln{x_1} (1 - \ln{\eta}) - \frac{2 \pi^2}{3} - 1\right ] H^{(0)},
\label{eq:g3a} \\
&&G^{(1)}_{3b} = -\frac{\alpha_s }{8 \pi N_c}\left [\frac{1}{\epsilon} +
\ln{\frac{4 \pi \mu^2}{x_1 \eta m^2_B e^{\gamma_E}}} - \frac{1}{2} \right ]H^{(0)},
\label{eq:g3b}\\
&&G^{(1)}_{3c} = -\frac{\alpha_s}{8 \pi N_c}\left [\frac{1}{\epsilon} +
\ln{\frac{4 \pi \mu^2}{\delta_{12} m^2_B e^{\gamma_E}}} -
\ln{\frac{\delta_2}{\delta_{12}}} \ln{\frac{\delta_1}{\delta_{12}}} -
\ln{\frac{\delta_1 \delta_2}{\delta^2_{12}}} - \frac{\pi^2}{3} \right ]H^{(0)},
\label{eq:g3c} \\
&&G^{(1)}_{3d} =  \frac{\alpha_s N_c}{8 \pi}\left [\frac{3}{\epsilon} +
3 \ln{\frac{4 \pi \mu^2}{\delta_{12} m^2_B e^{\gamma_E}}} -
\ln{\frac{\delta_1 \delta_2}{\delta^2_{12}}} + \frac{11}{2}
- \frac{2 \pi^2}{3} \right ]H^{(0)}, \label{eq:g3d}
\eeq
\beq
G^{(1)}_{3e} =  \frac{\alpha_s N_c}{8 \pi}\Bigl [\frac{3}{\epsilon} +
3 \ln{\frac{4 \pi \mu^2}{x_1 \eta  m^2_B e^{\gamma_E}}} -
\ln{\frac{\delta_2}{x_1 \eta}} (\ln{x_2} + 1)
 + \frac{1}{2} \ln{x_2} - \frac{\pi^2}{3} + \frac{7}{4} \Bigr ]H^{(0)}.
\label{eq:g3e}
\eeq
The amplitude $G^{(1)}_{3a}$ have no IR divergence due to the fact that
the radiative gluon attaches to the massive b quark and the internal
line in Fig.~\ref{fig:fig3}(a).
The amplitude $G^{(1)}_{3b}$ should have collinear divergence at the
first sight because the radiative gluon in Fig.~\ref{fig:fig3}(b)
attaches to the light valence quark, but it's found that the collinear
region $l \parallel p_2$ was suppressed, then $G^{(1)}_{3b}$ is IR finite.
The radiative gluon in Fig.~\ref{fig:fig3}(c) attaches to the light
valence anti-quarks, so that both the collinear and soft divergences
are produced in $G^{(1)}_{3c}$, where the large double logarithm
$\ln{\delta_1} \ln{\delta_2}$ denoted the overlap of the IR divergences
can be absorbed into the $B$ meson or the pion meson wave functions.
The radiative gluon in Fig.~\ref{fig:fig3}(d) attaches to the light
valence anti-quarks as well as the virtual LO hard gluon, so the soft
divergence and the large double logarithm aren't generated in $G^{(1)}_{3d}$.
The radiative gluon in Fig.~\ref{fig:fig3}(e) attaches only to the
light valence quark as well as the virtual LO hard gluon, and then
$G^{(1)}_{3e}$ just contains the collinear divergence regulated by
$\ln{\delta_2}$ from $l \parallel p_2$ region.

The analytical results from the box and pentagon diagrams as shown
in Fig.~\ref{fig:fig4} are summarized as
\beq
&&G^{(1)}_{4a} = -\frac{\alpha_s N_c}{8 \pi}\; x_1\; \left [ \ln{\frac{x_{2}
\eta^{2}}{\delta_{2}}} + 1 \right ] H^{(0)},
\label{eq:g4a}\\
&&G^{(1)}_{4b} = -\frac{\alpha_s C_f}{4 \pi}\left [\ln^{2}{\frac{\delta_1}{x^{2}_{1}}}
- \ln^{2}{x_{1}} - \frac{7 \pi^2}{3} \right ]H^{(0)},
\label{eq:g4b}\\
&&G^{(1)}_{4c} = -\frac{\alpha_s}{8 \pi  N_c}
\left [\ln{\frac{\delta_1}{\delta_{12}}} \ln{\frac{\delta_2}{\delta_{12}}}
- \frac{\pi^2}{12} \right ]H^{(0)}, \label{eq:g4c}\\
&&G^{(1)}_{4d} = -\frac{\alpha_s C_F}{2 \pi}
\left [\ln{\frac{\delta_1}{\delta_{12}}} \ln{\frac{\delta_2}{\delta_{12}}}
+ \frac{\pi^2}{3} \right ]H^{(0)}, \label{eq:g4d}\\
&&G^{(1)}_{4e} =  \frac{\alpha_s}{8 \pi  N_c}
\left [\ln{\frac{\delta_1}{\eta}} \ln{\frac{\delta_2}{\eta}} +
\ln{\delta_2 } + \frac{\pi^2}{6} \right ]H^{(0)}, \label{eq:g4e} \\
&&G^{(1)}_{4f} = -\frac{\alpha_s}{8 \pi  N_c}
\Bigl [\ln{\frac{\delta_1}{\eta}} \ln{\frac{\delta_2}{\eta}} -
\ln{x_{2}} \ln{\delta_{2}} - \ln{x_{2}} \ln{\delta_{12}} \non
&&~~~~~~~~~+ \frac{1}{2} \ln^{2}{\eta}
+ \frac{1}{2} \ln^{2}{\delta_{12}} - \frac{1}{2} \ln^{2}{x_{2}} -
\frac{\pi^2}{3} -1 \Bigr ]H^{(0)}. \label{eq:g4f}
\eeq
Note that the amplitude of Fig.~\ref{fig:fig4}(a) has no IR divergence
because the additional gluon is linked to the massive b quark and the
virtual LO hard kernel gluon.
Fig.~\ref{fig:fig4}(b) is two-particle reducible, whose IR contribution
would be canceled by the corresponding effective diagrams for the $B$
meson function Fig.~\ref{fig:fig5}(c).
All the other four subdiagrams Fig.~\ref{fig:fig5}(c,d,e,f) would
generate double logarithms from the overlap region of the soft and
collinear region, because the radiative gluon attached with b quark
and light valence quark generate both collinear divergence and soft
divergence, as well as the gluon attached two light valence partons.
Fig.~\ref{fig:fig4}(d) is also a two-particle reducible diagram,
whose contribution should be canceled completely by the corresponding
effective diagrams Fig.~\ref{fig:fig5}(c) for the pion meson function
due to the requirement of the factorization theorem.
It's found that the double logarithm in Fig.~\ref{fig:fig4}(c)
offset with the double logarithm in Fig.~\ref{fig:fig3}(c), and the
cancelation would also appear for the double logarithms in
Fig.~\ref{fig:fig4}(e) and Fig.~\ref{fig:fig4}(f).

The NLO twist-3 corrections from all the three kinds of
the QCD quark diagrams are summed into
\beq
G^{(1)} &=& \frac{\alpha_s C_f}{4 \pi}
\Biggl \{ \frac{21}{4} \left (\frac{1}{\epsilon} + \ln{\frac{4 \pi \mu^2}{m^2_B e^{\gamma_E}}} \right )
- \ln^{2}{\delta_{1}} - 2 \ln{\delta_1}\ln{\delta_2} - \frac{97}{16} \ln^2{x_1} - \frac{15}{8} \ln^2{x_2}
\non
&&+ \frac{1}{2} \left (-1 + 12 \ln{x_1} + 4 \ln{x_2} + 4 \ln{\eta} \right ) \ln{\delta_1}
+ \left (-1 + 2 \ln{x_1} + \ln{x_2} + 2 \ln{\eta} \right ) \ln{\delta_2} \non
&&
-\frac{23}{8} \ln{x_1} \ln{x_2}
-\frac{1}{8} \left (41 + 17 \ln{\eta} \right ) \ln{x_1}
- \frac{1}{16} \left (41 + 46 \ln{\eta} \right ) \ln{x_2} \non
&&-\frac{1}{96} \left [ -273 + \pi^2 + 96 \ln{r_g}
+ 12 \ln{\eta} \left (25 + 17 \ln{\eta} \right ) \right ] \Biggr \} H^{(0)},
\label{eq:nloqd}
\eeq
for $N_f = 6$. The UV divergence in the above expression is the same as in the
pion electromagnetic form factor\cite{prd83-054029} and in the leading twist
of $B \to \pi$ transition form factor\cite{prd85-074004}, which determines the
renormalization-group(RG) evolution of the the coupling constant $\alpha_s$.
The double logarithm arose from the reducible subdiagrams
Fig.~\ref{fig:fig4}(b,d) would be absorbed into the NLO wave functions.

\subsection{NLO Corrections of the Effective Diagrams}

As point out in Ref.~\cite{prd85-074004}, a basic argument of $k_T$ factorization
is that the IR divergences arisen from the NLO corrections can be absorbed into
the non-perturbative wave  functions which are universal.
From this point, the convolution of the NLO wave function $\Phi^{(1)}_B$
and the LO hard kernel $H^{(0)}$, the LO hard kernel $H^{(0)}$ and the NLO
wave function $\Phi^{(1)}_{\pi}$ are computed, and then to cancel the IR divergences
in the NLO amplitude $G^{(1)}$ as given in Eq.~(\ref{eq:nloqd}).
The convolutions for NLO wave functions and LO hard kernel are calculated
in this subsection.
In $k_T$ factorization theorem, the $\Phi^{(1)}_B$\cite{prd70-074030}
collect the $O(\alpha_s)$ effective diagrams from the matrix elements
of the leading Fock states $\Phi_B(x_1,k_{1T};x'_1,k'_{1T})$,
and $\Phi^{(1)}_{\pi,P}$ collect the $O(\alpha_s)$ effective
diagrams for the twist-3 transverse momenta dependent (TMD) light-cone
wave function $\Phi_{\pi,P}(x_2,k_{2T};x'_2,k'_{2T})$\cite{prd64-014019,epjc40-395}
\beq
&&\Phi_B(x_1,k_{1T};x'_1,k'_{1T}) = \int \frac{d z^-}{2 \pi}
\frac{d^2 z_T}{(2 \pi)^2} e^{-i x'_1 P^+_1 z^- + i \textbf{k}'_{1T}
\cdot \textbf{z}_T} \non
&&~~~~~~~~~~~~~~~~~~~~~~~\cdot <0\mid\overline{q}(z) W_z(n_1)^{\dag}
I_{n_1;z,0} W_0(n_1) \nsl_+ \Gamma h_{\nu}(0) \mid h_{\nu} \overline{d}(k_1)>,
\label{eq:nlowfb}
\eeq
\beq
&&\Phi_{\pi,P}(x_2,k_{2T};x'_2,k'_{2T}) = \int \frac{d y^+}{2 \pi}
\frac{d^2 y_T}{(2 \pi)^2} e^{-i x'_2 P^-_2 y^+ + i \textbf{k}'_{2T}
\cdot \textbf{y}_T} \non
&&~~~~~~~~~~~~~~~~~~~~~~~\cdot <0\mid\overline{q}(y) W_y(n_2)^{\dag}
I_{n_2;y,0} W_0(n_2) \gamma_5 q(0) \mid u(p_2 - k_2) \overline{d}(k_2)>,
\label{eq:nlowfpion}
\eeq
respectively, in which $z = (0, z_-, \textbf{z}_T)$ and
$y = (y^+, 0, \textbf{y}_T)$ are the light cone (LC) coordinates of the anti-quark
field $\overline{d}$ carried the momentum faction $x_i$ respectively,
and $h_{\nu}$ is the effective heavy-quark field.
\beq
&& W_z(n_1) = {\rm P}\; \exp \left [-i g_s \int^{\infty}_{0} d \lambda n_1
\cdot A(z + \lambda n_1) \right ],\\
&& W_y(n_2) = {\rm P}\; \exp \left [-i g_s \int^{\infty}_{0} d \lambda n_2 \cdot
A(y + \lambda n_2) \right ],
\label{eq:wilsonline}
\eeq
where ${\rm P}$ is the path ordering operator.
The two Wilson line $W_{y/z}(n_i)$ and $W_{0}(n_i)$ are connected by a vertical
link $I_{n_i;y/z,0}$ at infinity\cite{plb543-66}.
Then the additional LC singularities from the region where loop momentum
$l\parallel n_-(n_+) $\cite{appb34-3103} are regulated by the IR regulator
$n^2_1$ and $n^2_2$.
The scales $\xi^2_1 \equiv 4 (n_1 \cdot p_1)^2/ |n^2_1| = m^2_B |n^-_1/n^+_1|$
and $\xi^2_2 \equiv 4 (n_2 \cdot p_2)^2/ |n^2_2| = \eta^2 m^2_B |n^+_2/n^-_2|$
are introduced to avoid the LC singularity\cite{prd85-074004,jhep0601-067}.
It's important to emphasize that the variation of the above scales is regarded
as a factorization scheme-dependence,
which would be brought into the NLO hard kernel after taking the difference
between the QCD quark diagrams and the effective diagrams.
And the above scheme-dependent scales can be minimized by adhering to fixed
$n^2_1$ and $n^2_2$. In Ref.~\cite{jhep1401-004}, very recently, Li et al. studied
the joint resummation for pion wave function and pion transition form factor, i.e.,
summing up the mixed logarithm $\ln(x_i) \ln(\kt)$ to all orders. Such joint
resummation can reduce the above scheme dependence effectively.

\begin{figure}[tb]
\vspace{-2cm}
\begin{center}
\leftline{\epsfxsize=13cm\epsffile{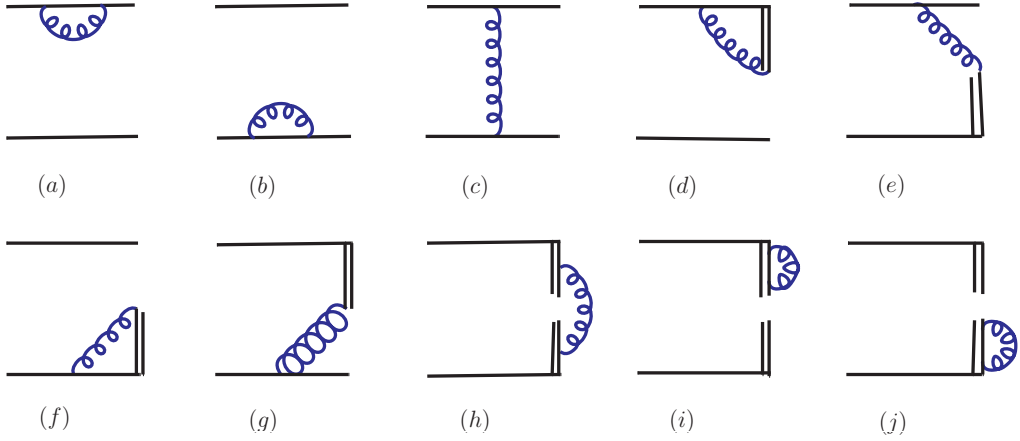}}
\end{center}
\vspace{-11cm}
\caption{$O(\alpha_s)$ diagrams for the B meson function.}
\label{fig:fig5}
\end{figure}

The convolution for $O(\alpha_s)$ order of $B$ meson function in
Eq.~(\ref{eq:nlowfb}) and $H^{(0)}$ over the integration variables
$x'_1$ and $k'_{1T}$ is
\beq
\Phi^{(1)}_B \otimes H^{(0)} \equiv \int dx'_1 d^2 \textbf{k}'_{1T}
\Phi^{(1)}_B(x_1,\textbf{k}_{1T};x'_1,\textbf{k}'_{1T}) H^{(0)}(x'_1,
\textbf{k}'_{1T};x_2,\textbf{k}_{2T}).
\label{eq:convob}
\eeq
In the evolution, the $n_1$ is approximated to vector $n_-$ with a very small plus
component $n^+_1$ to avoid the LC singularity in the integration,
and we choose $n^-_1$ to be positive while $n^+_1$ can be positive
or negative for convenience.
The NLO twist-3 corrections from the $O(\alpha_s)$ order wave function
as shown in Fig.~\ref{fig:fig5} are listed in the following with $\mu_f$
being the factorization scale:
\beq
&&\Phi^{(1)}_{5a} \otimes H^{(0)} = \frac{\alpha_s C_f}{4 \pi}
\left [\frac{1}{\epsilon} +
\ln{\frac{4 \pi \mu^2_f}{m^2_B e^{\gamma_E}}} - \ln{r_g} \right ] H^{(0)},
\label{eq:phi5a}\\
&&\Phi^{(1)}_{5b} \otimes H^{(0)} = -\frac{\alpha_s C_f}{8 \pi}
\left [\frac{1}{\epsilon} +
\ln{\frac{4 \pi \mu^2_f}{m^2_B e^{\gamma_E}}} - \ln{\delta_1} + 2 \right ] H^{(0)},
\label{eq:phi5b}\\
&&\Phi^{(1)}_{5c} \otimes H^{(0)} = -\frac{\alpha_s C_f}{4 \pi}
\left [\ln^{2}{(\frac{\delta_1}{x^{2}_{1}}}) \right ]H^{(0)},
\label{eq:phi5c}\\
&&\Phi^{(1)}_{5d} \otimes H^{(0)} = -\frac{\alpha_s C_f}{8 \pi}
(-\ln{r_1})\left [\frac{1}{\epsilon} +
\ln{\frac{4 \pi \mu^2_f}{m^2_B e^{\gamma_E}}} - \ln{r_g} \right ] H^{(0)},
\label{eq:phi5d}\\
&&\Phi^{(1)}_{5e} \otimes H^{(0)} =  \frac{\alpha_s C_f}{4 \pi} (-\ln{r_1})
\left [ - \ln{r_1} - \ln{r_g} + \frac{1}{2} \ln{r_1} + 2 \ln{x_1}\right ] H^{(0)},
\label{eq:phi5e}\\
&&\Phi^{(1)}_{5f} \otimes H^{(0)} =  \frac{\alpha_s C_f}{8 \pi}
\Biggl [ \frac{1}{\epsilon} +
\ln{\frac{4 \pi \mu^2_f}{m^2_B e^{\gamma_E}}} + \ln{r_1} - 2 \ln{x_1} \non
&&~~~~~~~~~~~- \left (\ln{\delta_1} - 2 \ln{x_1} + \ln{r_1} \right )^2
- 2 \left ( \ln{\delta_1} - 2 \ln{x_1} + \ln{r_1} \right )
- \frac{\pi^2}{3} + 2 \Biggr ] H^{(0)}, \label{eq:phi5f}\\
&&\Phi^{(1)}_{5g} \otimes H^{(0)} =  \frac{\alpha_s C_f}{8 \pi}
\left [ \left (\ln{\delta_1} -
2 \ln{x_1} + \ln{r_1} \right )^2 - \frac{\pi^2}{3} \right ] H^{(0)},
\label{eq:phi5g}\non
&&\left (\Phi^{(1)}_{5h} + \Phi^{(1)}_{5i} + \Phi^{(1)}_{5j} \right )
\otimes H^{(0)} = \frac{\alpha_s C_f}{4 \pi}
\left [\frac{1}{\epsilon} +
\ln{\frac{4 \pi \mu^2_f}{m^2_B e^{\gamma_E}}} - \ln{\delta_{12}} \right ] H^{(0)},
\label{eq:phi5sum}
\eeq
where the dimensionless parameter $r_1 = m^2_B / \xi^2_1$ is chosen small
to obtain the simple results as above.
Because the two propagators in the LO hard kernel $H^{(0)}$ are both
relevant to $x'_1$ while only one is relevant to $x'_2$,
there exist three 5-point integrals as shown in Fig.~\ref{fig:fig5}(c,e,g)
need to be calculated.
The reducible subdiagrams Fig.~\ref{fig:fig5}(c) reproduced the double
logarithm as the quark subdiagram Fig.~\ref{fig:fig4}(b).
Difference between the effective heavy-quark field employed in the $B$
meson wave function and the b quark field in the quark diagrams leads
to different results in Fig.~\ref{fig:fig4}(b) and Fig.~\ref{fig:fig5}(c).
It's found that the regulator $\ln{m_g}$ adopted to regularize the soft
divergence in the reducible Fig.~\ref{fig:fig5}(a) will be canceled by
the Fig.~\ref{fig:fig2}(a), while the regulators $\ln{m_g}$ in
Fig.~\ref{fig:fig5}(d) and Fig.~\ref{fig:fig5}(e) cancels each other.
The large double logarithms $(\ln\delta_1 - 2 \ln{x_1}
+ \ln{r_1} )^2$ in Fig.~\ref{fig:fig5}(f) and Fig.~\ref{fig:fig5}(g)
also cancel each other.
So the other IR divergences are regulated only by $\ln{\delta_1}$ as
the prediction because it's just the NLO correction to the incoming
$B$ meson wave function.

After summing all the $O(\alpha_s)$ contributions in Fig.~\ref{fig:fig5},
we obtain
\beq
\Phi^{(1)}_B \otimes H^{(0)} = &&\frac{\alpha_s C_f}{4 \pi}
[\frac{1}{2}(4 + \ln{r_1})(\frac{1}{\epsilon} + \ln{\frac{4 \pi
\mu^2_f}{m^2_B e^{\gamma_E}}})
- \ln^{2}{\delta_{1}} \non
&&+ \frac{1}{2} (-1 + 8 \ln{x_{1}}) \ln{\delta_1} - \ln{r_g}
- 4 \ln^{2}{x_{1}} + (1 - \ln{r_1}) \ln{x_1}\non
&& - \frac{1}{2} \ln{r_{1}} + \frac{1}{4} \ln^{2}{r_{1}}
- \ln{\delta_{12}} -\frac{\pi^{2}}{3}] H^{(0)}.
\label{eq:nloedtb}
\eeq

The convolution of $H^{(0)}$ and the $O(\alpha_s)$ outgoing
pion meson wave function $\Phi^{(1)}_{\pi}$ over the integration
variables $x'_2$ and $k'_{2T}$ is
\beq
H^{(0)} \otimes \Phi^{(1)}_{\pi,P} \equiv \int dx'_2 d^2 \textbf{k}'_{2T}
H^{(0)}(x_1,\textbf{k}_{1T};x'_2,\textbf{k}'_{2T}) \Phi^{(1)}_{\pi,P}(x'_2,
\textbf{k}'_{2T};x_2,\textbf{k}_{2T}) .
\label{eq:convopi}
\eeq
The $n_2$ is mainly in $n_+$ component, and a very small minus component
$n^-_2$ is kept to avoid the LC singularity.
Note that the sign of $n^+_2$ is positive as $P^+_1$ while the sign of
$n^-_2$ is arbitrary for convenience.
\begin{figure}[tb]
\vspace{-2cm}
\begin{center}
\leftline{\epsfxsize=13cm\epsffile{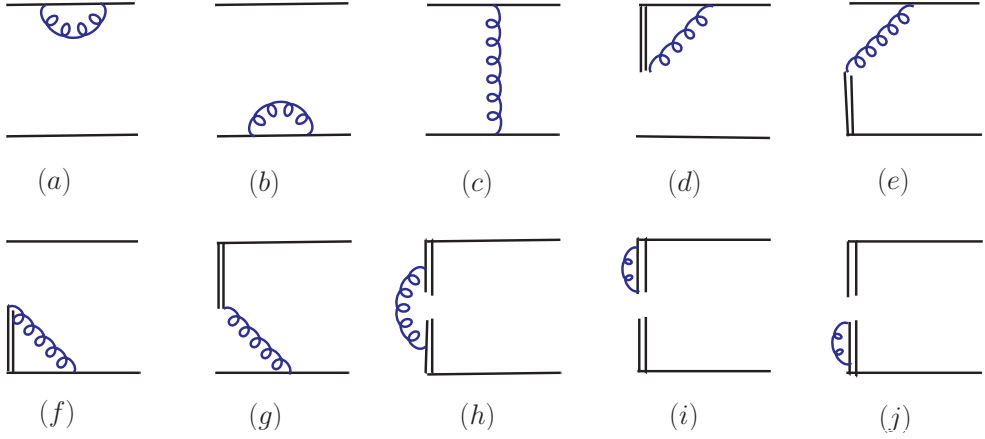}}
\end{center}
\vspace{-12cm}
\caption{$O(\alpha_s)$ diagrams for the $\pi$ meson function.}
\label{fig:fig6}
\end{figure}

Fig.~\ref{fig:fig6} collects all the NLO corrections to the outgoing
pion wave function, and $r_2 = m^2_B / \xi^2_2$.
The reducible subdiagrams Fig.~\ref{fig:fig5}(a,b,c) and
Fig.~\ref{fig:fig6}(a,b,c) generate the same results as in the leading
twist-2 case\cite{prd85-074004},
while the results from the inreducible subdiagrams
Fig.~\ref{fig:fig5}(d,e,f,g,h,i,j) and Fig.~\ref{fig:fig6}(d,e,f,g,h,i,j)
in the twist-3 is half smaller than that in the leading twist-2,
due to their different spin structures.
The amplitude of the reducible Fig.~\ref{fig:fig6}(c), convoluted by the
LO hard kernel $H^{(0)}$, reproduced the double logarithm
$\ln{\delta_1} \ln{\delta_2}$.
There are no five-point integrals in $H^{(0)} \otimes \Phi^{(1)}$ because
only one denominator in $H^{(0)}$ is relevant to $x'_{2}$.
Then the most complicated integrals involved here is the four-point integrations
attached to Fig.~\ref{fig:fig6}(e,g).
The double logarithm in $H^{(0)} \otimes \Phi^{(1)}_{8d}$, $H^{(0)}
\otimes \Phi^{(1)}_{8e}$, $H^{(0)} \otimes \Phi^{(1)}_{8f}$ and $H^{(0)}
\otimes \Phi^{(1)}_{8g}$ are also canceled. Only a double logarithm
$\ln{\delta_1} \ln{\delta_2}$, which would be canceled by the quark
diagram Fig.~\ref{fig:fig4}(d), still left in the $H^{(0)} \otimes \Phi^{(1)}_{\pi,P}$.

The analytical results from Fig.~\ref{fig:fig6} are listed in the
following with $\mu_f$ being the factorization scale.
\beq
&&H^{(0)} \otimes \Phi^{(1)}_{6a} = -\frac{\alpha_s C_f}{8 \pi}
\left [\frac{1}{\epsilon} + \ln{\frac{4 \pi \mu^2_f}{m^2_B e^{\gamma_E}}}
- \ln{\delta_2} + 2 \right ] H^{(0)}, \label{eq:phi6a}\\
&&H^{(0)} \otimes \Phi^{(1)}_{6b} = -\frac{\alpha_s C_f}{8 \pi}
\left [\frac{1}{\epsilon} + \ln{\frac{4 \pi \mu^2_f}{m^2_B e^{\gamma_E}}}
- \ln{\delta_2} + 2 \right ] H^{(0)}, \label{eq:phi6b}\\
&&H^{(0)} \otimes \Phi^{(1)}_{6c} = -\frac{\alpha_s C_f}{2 \pi}
\left [\ln{\frac{\delta_{12}}{\delta_1}} \ln{\frac{\delta_{12}}{\delta_2}}
+ \frac{\pi^2}{3} \right ] H^{(0)}, \label{eq:phi6c}\\
&&H^{(0)} \otimes \Phi^{(1)}_{6d} =  \frac{\alpha_s C_f}{8 \pi}
\Biggl [\frac{1}{\epsilon} + \ln{\frac{4 \pi \mu^2_f}{m^2_B e^{\gamma_E}}}
- \ln{\delta_2} - \left (\ln{r_2} + \ln{\delta_2} \right )^2 \non
&&~~~~~~~~~~~~~~~~~~~~~~~~~- \left (\ln{r_2} + \ln{\delta_2} \right )
+ 2 - \frac{\pi^2}{3} \Biggr] H^{(0)}, \label{eq:phi6d}\\
&&H^{(0)} \otimes \Phi^{(1)}_{6e} =  \frac{\alpha_s C_f}{8 \pi}
\Bigl [ \left (\ln{x_2} - \ln{r_2} - \ln{\delta_2} \right )^2 + \pi^2 \Bigr ] H^{(0)},
\label{eq:phi6e}
\eeq
\beq
&& H^{(0)} \otimes \Phi^{(1)}_{6f} =  \frac{\alpha_s C_f}{8 \pi}
\Biggl [\frac{1}{\epsilon} + \ln{\frac{4 \pi \mu^2_f}{m^2_B e^{\gamma_E}}}
- \ln{\delta_2} - \left (2 \ln{x_2} - \ln{r_2} - \ln{\delta_2} \right )^2 \non
&& ~~~~~~~~~~~~~~~~~~~~~~~~~+ \left (2 \ln{x_2} - \ln{r_2} - \ln{\delta_2}\right )
+ 2 - \frac{\pi^2}{3} \Biggr ] H^{(0)},
\label{eq:phi6f} \\
&&H^{(0)} \otimes \Phi^{(1)}_{6g} =  \frac{\alpha_s C_f}{8 \pi}
\left [ \left (2 \ln{x_2} - \ln{r_2} - \ln{\delta_2} \right )^2 - \frac{\pi^2}{3}\right ]
H^{(0)}, \label{eq:phi6g}\\
&& H^{(0)} \otimes \left (\Phi^{(1)}_{6h} + \Phi^{(1)}_{6i} + \Phi^{(1)}_{6j} \right)
=  \frac{\alpha_s C_f}{4 \pi} \left [\frac{1}{\epsilon} +
\ln{\frac{4 \pi \mu^2_f}{m^2_B e^{\gamma_E}}} - \ln{\delta_{12}} \right ] H^{(0)}.
\label{eq:phi6sum}
\eeq

The total contributions from the convolution of the LO hard kernel and
the NLO final pion meson wave function is obtained  by summing all
terms as given in above equations.
\beq
H^{(0)} \otimes \Phi^{(1)}_{\pi,P} &=&\frac{\alpha_s C_f}{4 \pi}
\Biggl [ \left (\frac{1}{\epsilon} + \ln{\frac{4 \pi \mu^2_f}{m^2_B e^{\gamma_E}}}\right )
- 2 \ln{\delta_1} \ln{\delta_2} + 2 \ln{\delta_{12}} \ln{\delta_1} \non
&&- \left ( 1 + \ln{x_2} - 2 \ln{\delta_{12}} \right ) \ln{\delta_2}
+ \frac{1}{2}\ln^2{(x_2)} + (1 - \ln{r_2}) \ln{x_2} \non
&&- \ln{r_2} - \ln{\delta_{12}} -2 \ln^{2}{\delta_{12}}
- \frac{2 \pi^{2}}{3} \Biggr ] H^{(0)}.
\label{eq:nloedtpi}
\eeq

\subsection{NLO Hard Kernel}

It's obvious that the UV poles are different in Eq.~(\ref{eq:nloedtb})
and Eq.~(\ref{eq:nloedtpi}), since the former involves the effective
heavy-quark field, instead of the b quark field. Then the $B$ meson
and pion meson wave functions exhibit different evolution as
proved in Ref.~\cite{prd85-074004}.
The $\ln{\mu_f}$ term in Eq.~(\ref{eq:nloedtb}) was partly absorbed into
the $B$ meson wave function, and partly to the $B$ meson decay constant $f_B(\mu_f)$.

The IR-finite $\kt$ dependent NLO hard kernel for the $B \to \pi$
transition form factor at twist-3 is extracted by taking the difference
between the contributions from QCD quark diagrams and the contributions
from effective diagrams \cite{prd67-034001}.
\beq
&&H^{(1)}(x_1,\textbf{k}_{1T};x_2,\textbf{k}_{2T}) = G^{(1)}
(x_1,\textbf{k}_{1T};x_2,\textbf{k}_{2T}) \non
&&~~~~~~~~~~~~~~- \int dx'_1 d^2 \textbf{k}'_{1T}
\Phi^{(1)}_B(x_1,\textbf{k}_{1T};x'_1,\textbf{k}'_{1T})
H^{(0)}(x'_1,\textbf{k}'_{1T};x_2,\textbf{k}_{2T}) \non
&&~~~~~~~~~~~~~~- \int dx'_2 d^2 \textbf{k}'_{2T}
H^{(0)}(x_1,\textbf{k}_{1T};x'_2,\textbf{k}'_{2T})
\Phi^{(1)}_{\pi,P}(x'_2,\textbf{k}'_{2T};x_2,\textbf{k}_{2T}).
\label{eq:nlohk}
\eeq
The bare coupling constant $\alpha_s$ in
Eq.~(\ref{eq:nloqd},\ref{eq:nloedtb},\ref{eq:nloedtpi}) can be rewritten as
\beq
&&\alpha_s = \alpha_s(\mu_f) + \delta Z(\mu_f) \alpha_s(\mu_f),
\label{eq:renormcc}
\eeq
in which the counterterm $\delta Z(\mu_f)$ is
defined in the $\overline{MS}$ scheme.
Inserting Eq.~(\ref{eq:renormcc})
into Eqs.~(\ref{eq:lohk},\ref{eq:nloqd}) and
(\ref{eq:nloedtb},\ref{eq:nloedtpi})
regularizes the UV poles in Eq.~(\ref{eq:nlohk}) through the multiplication
$\delta Z (\mu_f) H^{(0)}$, and then the UV poles in
Eqs.~(\ref{eq:nloedtb},\ref{eq:nloedtpi}) are regulated by the
counterterm of the quark field and by an additional counterterm in the
$\overline{MS}$ scheme.

The NLO hard kernel $H^{(1)}$ for Fig.~\ref{fig:fig1}(b) at twist-3 is given by
\beq
H^{(1)} = &&\frac{\alpha_s(\mu_f) C_F}{4 \pi}
\Biggl\{ \frac{21}{4} \ln{\frac{\mu^2}{m^2_B}}
- \frac{1}{2}(6 + \ln{r_1}) \ln{\frac{\mu^2_f}{m^2_B}}
- \frac{1}{16} \ln^2{x_1} - \frac{3}{8} \ln^2{x_2} \non
&&+ \frac{9}{8} \ln{x_1} \ln{x_2} + \left (- \frac{33}{8}
+ \ln{r_1} + \frac{15}{8} \ln{\eta} \right ) \ln{x_1}
+ \left (- \frac{25}{16} + \ln{r_2} + \frac{9}{8} \ln{\eta} \right) \ln{x_2} \non
&&+ \frac{1}{2} \ln{r_1} - \frac{1}{4} \ln^{2}{r_1}
+ \ln{r_2} - \frac{9}{8} \ln{\eta} - \frac{1}{8} \ln^{2}{\eta}
+ \frac{95 \pi^2}{96} + \frac{273}{96} \Biggr \} H^{(0)}.
\label{eq:nlohk1}
\eeq
The choice of the dimensionless scales $\xi^2_1$ and $\xi^2_2$
corresponds to a factorization scheme as discussed in the last subsection.
The $\xi^2_2$ is fixed to $m^2_B$ and $\xi_1 / m_B = 25$
is chosen in the numerical analysis to obtain the simplified
results in Eqs.~(39-42).

The additional double logarithm $\ln^{2}{x_1}$ derived from the
limit that the internal quark is on-shell due to the tiny momentum
fraction $x_1$ should be considered.
It's absorbed into the jet function $J(x_1)$ \cite{prd66-094010,plb555-197}
\beq
J^{(1)} H^{(0)} = &&- \frac{1}{2} \frac{\alpha_s(\mu_f) C_F}{4 \pi}
\left [\ln^2{(x_1)} + \ln{x_1} + \frac{\pi^2}{3} \right ] H^{(0)},
\label{eq:jetfunction}
\eeq
where the factor $1/2$ reflects the different spin structures
of the twist-3 and twist-2 cases.
The NLO hard kernel from Eq.~(\ref{eq:nlohk1}) turns into the following
format after subtracting the jet function in Eq.~(\ref{eq:jetfunction})
\beq
H^{(1)} \to &&H^{(1)} - J^{(1)} H^{(0)} \non
&=&\frac{\alpha_s(\mu_f) C_F}{4 \pi}
\Biggl [\frac{21}{4} \ln{\frac{\mu^2}{m^2_B}}
- \frac{1}{2}(6 + \ln{r_1}) \ln{\frac{\mu^2_f}{m^2_B}}
+ \frac{7}{16} \ln^2{x_1} - \frac{3}{8} \ln^2{x_2} \non
&&+ \frac{9}{8} \ln{x_1} \ln{x_2}
+ \left (- \frac{29}{8}+ \ln{r_1} + \frac{15}{8} \ln{\eta} \right ) \ln{x_1}
+ \left (- \frac{25}{16} + \ln{r_2} + \frac{9}{8} \ln{\eta} \right) \ln{x_2} \non
&&+ \frac{1}{2} \ln{r_1} - \frac{1}{4} \ln^{2}{r_1} + \ln{r_2}
- \frac{9}{8} \ln{\eta} - \frac{1}{8} \ln^{2}{\eta} + \frac{37 \pi^2}{32}
+ \frac{91}{32} \Biggr ] H^{(0)}\non
&=& F^{(1)}_{\rm T3}(x_i,\mu,\mu_f,q^2)\; H^{(0)},
\label{eq:nlohk2}
\eeq
where $r_i=m_B^2/\xi_i^2$, $\eta=1-q^2/m_B^2$.
The IR-finite $\kt$ dependent function $F^{(1)}_{\rm T3}(x_i,\mu,\mu_f,q^2)$ in
Eq.~(\ref{eq:nlohk2}) describes the NLO twist-3 contribution to the
$B \to \pi$ transition form factor $f^+(q^2)$ and $f^0(q^2)$ as defined in
Eq.~(\ref{eq:ffme}).

\section{Numerical Analysis}

In this section, the $B \to \pi$ transition form factors will be evaluated numerically
up to twist-3 by employing the $\kt$ factorization theorem,
and the comparative analysis is developed between the LO and NLO as
well as between the twist-2 and twist-3 NLO corrections.

In the calculation, the following non-asymptotic pion  distribution amplitudes (DAs)
as given in Refs.~\cite{zpc48-239,jhep01-010} will be used:
\beq
\phi_{\pi}^{A}(x) &=&\frac{3 f_{\pi}}{ \sqrt{6}} x (1-x) ~
\Bigl [1 + a_2^{\pi} C_2^{\frac{3}{2}}(u)
                    + a_4^{\pi} C_4^{\frac{3}{2}}(u) \Bigr ], \non
\phi_{\pi}^{P}(x) &=&\frac{f_{\pi}}{2 \sqrt{6}}
\Bigl [ 1 + \left ( 30 \eta_3 - \frac{5}{2} \rho_{\pi}^2 \right) C_2^{\frac{1}{2}}(u)
 - 3 \left (\eta_3 \omega_3 + \frac{9}{20} \rho_{\pi}^2
                    (1 + 6 a_2^{\pi}) \right ) C_4^{\frac{1}{2}}(u) \Bigr ], \non
\phi_{\pi}^{T}(x) &=&\frac{f_{\pi}}{2 \sqrt{6}} (1-2 x)
~\left [ 1 + 6 \left (5 \eta_3 - \frac{1}{2} \eta_3 \omega_3
                    - \frac{7}{20} \rho_{\pi}^2 - \frac{3}{5}
                    \rho_{\pi}^2 a_2^{\pi} \right )
                    \left (1-10 x + 10 x^2 \right ) \right ],
\label{eq:dapion}
\eeq
where $u=2x-1$, $m_{\pi} = 0.135$ GeV, $f_{\pi} = 0.13$ GeV, $m_0^{\pi} = 1.4$ GeV,
and the Gegenbauer moments and Gegenbauer polinomials are adopted from
Refs.~\cite{prd71-014015,jhep0605-004},
\beq
a_2^{\pi} &=& 0.25, \quad a_4^{\pi} = -0.015, \rho_{\pi} = \frac{m_{\pi}}{m_0^{\pi}},
\quad \eta_3 = 0.015, \quad \omega_3 = -3.0, \label{eq:inputs}\\
C_2^{1/2} &=& \frac{1}{2}\left (3 u^2 - 1 \right ),
\quad C_2^{3/2} = \frac{3}{2} \left (5 u^2 - 1 \right ), \non
C_4^{1/2} &=& \frac{1}{8} \left ( 3 - 30 u^2 +35 u^4 \right ),
\quad C_4^{3/2} = \frac{15}{8} \left (1 - 14 u^2 + 21 u^4 \right ),
\label{eq:newgg}
\eeq

The $B$ meson distribution amplitude widely used in the
pQCD approach is of the form \cite{prd65-014007,epjc23-275}
\beq
\phi_B(x,b) &=& \frac{f_B}{2 \sqrt{6}} N_B x^2 (1-x)^2
\cdot \exp\left [-\frac{x^2 m_B^2}{2 \omega_B^2} - \frac{1}{2} (\omega_B b)^2 \right ],
\label{eq:daB}
\eeq
where we have assumed that $\phi_B(x,b) =\phi^+_B(x,b)=\phi^-_B(x,b)$.
The normalization condition of $\phi_B(x,b)$ is
\beq
\int_0^1{dx \phi_B(x,~b=0)} = \frac{f_B}{2 \sqrt{2 N_c}},\label{eq:ncB}
\eeq
with the mass $m_B = 5.28$ GeV,
while the normalization constant $N_B = 100.921$ for $f_B=0.21$ GeV and
the fixed shape parameter $\omega_B = 0.40$.

The form factor $f^+(q^2)$ and $f^0(q^2)$ at full LO level can be written as
\cite{prd65-014007}
\beq
f^+(q^2)|_{\rm LO} &=& 8 \pi m^2_B C_F \int{dx_1 dx_2} \int{b_1 db_1 b_2 db_2} \phi_B(x_1,b_1)
\non
&& \times \Biggl \{ r_\pi  \left [\phi_{\pi}^{P}(x_2) - \phi_{\pi}^{T}(x_2) \right ]
\cdot \alpha_s(t_1)\cdot e^{-S_{B\pi}(t_1)}\cdot S_t(x_2)\cdot
h(x_1,x_2,b_1,b_2) \non
&& + \left [ (1 + x_2 \eta) \phi_{\pi}^A(x_2)+ 2 r_\pi \left ( \frac{1}{\eta} - x_2 \right )
\phi_{\pi}^T(x_2) - 2x_2 r_\pi \phi_{\pi}^P(x_2) \right ] \non
&& \hspace{1cm}\cdot \alpha_s(t_1)\cdot e^{-S_{B\pi}(t_1)} \cdot
S_t(x_2)\cdot h(x_1,x_2,b_1,b_2)\non
&& + 2 r_{\pi} \phi_{\pi}^P(x_2)\cdot \alpha_s(t_2)\cdot e^{-S_{B\pi}(t_2)}
\cdot S_t(x_1)\cdot h(x_2,x_1,b_2,b_1) \Biggr \}, \label{eq:ff11}
\eeq
\beq
f^0(q^2)|_{\rm LO} &=& 8 \pi m^2_B C_F \int{dx_1 dx_2} \int{b_1 db_1 b_2 db_2} \phi_B(x_1,b_1)
\non
&& \times \Biggl \{ r_\pi (2-\eta)\left [\phi_{\pi}^{P}(x_2) - \phi_{\pi}^{T}(x_2) \right ]
\cdot \alpha_s(t_1)\cdot e^{-S_{B\pi}(t_1)}\cdot S_t(x_2)\cdot
h(x_1,x_2,b_1,b_2) \non
&& + \left [ (1 + x_2 \eta)\eta \phi_{\pi}^A(x_2)
+ 2 r_\pi \left ( 1 - x_2\eta \right )
\phi_{\pi}^T(x_2) - 2x_2 \eta r_\pi \phi_{\pi}^P(x_2) \right ] \non
&& \hspace{1cm}\cdot \alpha_s(t_1)\cdot e^{-S_{B\pi}(t_1)} \cdot
S_t(x_2)\cdot h(x_1,x_2,b_1,b_2)\non
&& + 2 \eta r_{\pi} \phi_{\pi}^P(x_2)\cdot \alpha_s(t_2)\cdot e^{-S_{B\pi}(t_2)}
\cdot S_t(x_1)\cdot h(x_2,x_1,b_2,b_1) \Biggr\},
\label{eq:ff10}
\eeq
where $r_{\pi} = m_0^{\pi}/m_B$, the term proportional to $\phi_\pi^A$ denotes the LO twist-2
contribution, while those proportional to $\phi_\pi^P$ and $\phi_\pi^T$ make up of
the LO twist-3 contribution. The factor $\exp[-S_{B\pi}(t)]$ in
Eqs.~(\ref{eq:ff11},\ref{eq:ff10}) contains the
Sudakov logarithmic corrections and the renormalization group evolution effects
of both the wave functions and the hard scattering amplitude with
$S_{B\pi}(t)=S_B(t)+S_\pi(t)$, where
\beq
S_B(t)&=&s\left (x_1\frac{m_B}{\sqrt{2}},b_1 \right )+\frac{5}{3}\int_{1/b_1}^{t}
\frac{d\bar{\mu}}{\bar{\mu}}\gamma_q\left (\alpha_s(\bar{\mu}) \right ),\non
S_\pi(t)&=&s\left (x_2\frac{m_B}{\sqrt{2}},b_2 \right )
+s\left ((1-x_2)\frac{m_B}{\sqrt{2}},b_2 \right )
+2\int_{1/b_2}^{t}\frac{d\bar{\mu}}{\bar{\mu}}\gamma_q\left (\alpha_s(\bar{\mu}) \right ),
\eeq
with the quark anomalous dimension $\gamma_q=-\alpha_s/\pi$.
The functions $s(Q,b)$ are defined by \cite{prd65-014007}
\beq
s(Q,b)&=&\frac{A^{(1)}}{2\beta_1}\hat{q}\ln\left (\frac{\hat{q}}{\hat{b}}\right)
-\frac{A^{(1)}}{2\beta_1}(\hat{q}-\hat{b})
+\frac{A^{(2)}}{4\beta_1^2}\left (\frac{\hat{q}}{\hat{b}}-1 \right )\non
&&-\left [\frac{A^{(2)}}{4\beta_1^2}-\frac{A^{(1)}}{4\beta_1}
\ln\left (\frac{e^{2\gamma_E}-1}{2}\right )\right ]
\ln\left ({\frac{\hat{q}}{\hat{b}}}\right )\non
&& +\frac{A^{(1)}\beta_2}{4\beta_1^3}\hat{q}\left [\frac{\ln(2\hat{q})+1}{\hat{q}}
-\frac{\ln(2\hat{b})+1}{\hat{b}} \right ]
+\frac{A^{(1)}\beta_2}{8\beta_1^3} \left [ \ln^2(2\hat{q})-\ln^2(2\hat{b})\right ],
\eeq
where the variables are defined by $ \hat{q}=\ln[Q/(\sqrt{2}\Lambda)]$,
$\hat{b}=\ln[1/(b\Lambda)]$, and the coefficients $A^{(i)}$ and $\beta_i$ are
\beq
\beta_1&=&\frac{33-2n_f}{12}, \quad \beta_2=\frac{153-19n_f}{24},\quad
A^{(1)}=\frac{4}{3}, \notag \\
 \quad A^{(2)}&=&\frac{67}{9}-\frac{\pi^2}{3}-\frac{10n_f}{27}
+\frac{8}{3}\beta_1\ln(e^{\gamma_E}/2).
\eeq
Here, $n_f$ is the number of the quark flavors, and the $\gamma_E$ is the Euler constant.
The hard scales $t_i$ in the equations of this work are chosen as the largest
scale of the virtuality of the internal particles in the hard
$b$-quark decay diagram,
\beq
t_1=\max\{\sqrt{x_2\eta}m_B,1/b_1,1/b_2\},  \quad
t_2=\max\{\sqrt{x_1\eta}m_B,1/b_1,1/b_2\}. \label{eq:maxti}
\eeq

The function $S_t(x)$ in Eqs.~(\ref{eq:ff11},\ref{eq:ff10}) is the
threshold resummation factor that is adopted from Ref.~\cite{prd65-014007}:
\beq
S_t(x)=\frac{2^{1+2c}\Gamma(3/2+c)}{\sqrt{\pi}\Gamma(1+c)}[x(1-x)]^c,
\eeq
where we set the  parameter $c=0.3$.
The hard functions $h(x_1,x_2,b_1,b_2)$ come from the Fourier transform of the hard
kernel and can be written as \cite{prd83-054029}
\beq
h(x_1,x_2,b_1,b_2)&=&K_0\left (\sqrt{x_1x_2\eta}m_Bb_1 \right )
\Bigl [\theta(b_1-b_2)I_0\left (\sqrt{x_2\eta}m_Bb_2\right )
K_0\left (\sqrt{x_2\eta}m_Bb_1 \right)\non
&& +\theta(b_2-b_1)I_0\left (\sqrt{x_2\eta}m_Bb_1 \right )
K_0\left (\sqrt{x_2\eta}m_Bb_2\right )\Bigr],
\eeq
where $I_0$ and $K_0$ are the modified Bessel functions.

Before taking the NLO twist-3 contributions into account, we have to
make a choice for the scale $\mu$ and $\mu_f$, and try to minimize the
NLO correction to the form factor.
Following Ref.~\cite{prd85-074004}, we also set $\mu_f=t$ with $t=t_1$ or $t_2$,
the hard scale specified in the pQCD approach as given in
Eq.~(\ref{eq:maxti}), which is the largest energy scale in Fig.~\ref{fig:fig1}(a)
or ~\ref{fig:fig1}(b) respectively.
The renormalization scale $\mu$ is chosen to diminish all the single-logarithm
and constant terms in the NLO hard kernel (\ref{eq:nlohk2}) \cite{prd85-074004}:
\beq
t_s(\mu_f)= \left \{ {\rm Exp } \left [ c1+ \left ( -\frac{9}{4}
+\frac{1}{2}\ln{r_1} \right )
\ln{ \frac{\mu^2_f}{m^2_B} }
x^{c2}_1 x^{c3}_2 \right ] \right \}^{2/21}\cdot \mu_f,
\label{eq:scaleren}
\eeq
with the coefficients
\beq
&&c1=- \left ( \frac{1}{2}-\frac{1}{4}\ln{r_1} \right) \ln{r_1}
+ \left ( \frac{9}{8}+\frac{1}{8}\ln{\eta} \right ) \ln{\eta}
- \frac{379}{32} - \frac{167 \pi^2}{96},\non
&& c2= \frac{29}{8} - \ln{r_1} - \frac{15}{8}\ln{\eta},\non
&& c3= \frac{25}{16} - \frac{9}{8}\ln{\eta},
\label{eq:scalerencoeff}
\eeq
based on our calculation.

When the NLO twist-2 and NLO twist-3 contributions
to the $B \to \pi$ transition form factors are taken into account, the pQCD predictions for
the two form factors at full NLO level are of the form
\beq
f^+(q^2)|_{\rm NLO} &=& 8 \pi m^2_B C_F \int{dx_1 dx_2} \int{b_1 db_1 b_2 db_2} \phi_B(x_1,b_1)\non
&& \hspace{-1cm}\times \Biggl \{ r_\pi  \left [\phi_{\pi}^{P}(x_2) - \phi_{\pi}^{T}(x_2) \right ]
\cdot \alpha_s(t_1)\cdot e^{-S_{B\pi}(t_1)}\cdot S_t(x_2)\cdot
h(x_1,x_2,b_1,b_2) \non
&& \hspace{-1cm} + \left [ (1 + x_2 \eta)
\left (1 + F^{(1)}_{\rm T2}(x_i,t,q^2)\; \right )
\phi_{\pi}^A(x_2)+ 2 r_\pi \left ( \frac{1}{\eta} - x_2 \right )
\phi_{\pi}^T(x_2) - 2x_2 r_\pi \phi_{\pi}^P(x_2) \right ] \non
&& \cdot \alpha_s(t_1)\cdot e^{-S_{B\pi}(t_1)} \cdot
S_t(x_2)\cdot h(x_1,x_2,b_1,b_2)\non
&& \hspace{-1cm} + 2 r_{\pi} \phi_{\pi}^P(x_2)
\left (1 + F^{(1)}_{\rm T3}(x_i,t,q^2)\; \right )
\cdot \alpha_s(t_2)\cdot e^{-S_{B\pi}(t_2)}
\cdot S_t(x_1)\cdot h(x_2,x_1,b_2,b_1) \Biggr \}, \label{eq:ffnlop}
\eeq
\beq
f^0(q^2)|_{\rm NLO} &=& 8 \pi m^2_B C_F \int{dx_1 dx_2} \int{b_1 db_1 b_2 db_2} \phi_B(x_1,b_1)
\non
&& \hspace{-1cm} \times \Biggl \{ r_\pi (2-\eta)\left [\phi_{\pi}^{P}(x_2) - \phi_{\pi}^{T}(x_2) \right ]
\cdot \alpha_s(t_1)\cdot e^{-S_{B\pi}(t_1)}\cdot S_t(x_2)\cdot
h(x_1,x_2,b_1,b_2) \non
&& \hspace{-1cm} + \left [ (1 + x_2 \eta)
\left (1 + F^{(1)}_{\rm T2}(x_i,t,q^2)\right ) \eta \phi_{\pi}^A(x_2)
+ 2 r_\pi \left ( 1 - x_2\eta \right )
\phi_{\pi}^T(x_2) - 2x_2 \eta r_\pi \phi_{\pi}^P(x_2) \right ] \non
&&\cdot \alpha_s(t_1)\cdot e^{-S_{B\pi}(t_1)} \cdot
S_t(x_2)\cdot h(x_1,x_2,b_1,b_2)\non
&&\hspace{-1cm} + 2 \eta r_{\pi}
\left (1 + F^{(1)}_{\rm T3}(x_i,t,q^2) \right ) \phi_{\pi}^P(x_2)\cdot \alpha_s(t_2)\cdot e^{-S_{B\pi}(t_2)}
\cdot S_t(x_1)\cdot h(x_2,x_1,b_2,b_1) \Biggr\},
\label{eq:ffnlo0}
\eeq
where the factor $F^{(1)}_{\rm T2}(x_i,t,q^2)$ describes the NLO twist-2 contribution
as given in Ref.~\cite{prd85-074004}
\beq
F^{(1)}_{\rm T2}(x_i,t,q^2)&=& \frac{\alpha_s(\mu_f) C_F}{4 \pi}
\Biggl [\frac{21}{4} \ln{\frac{\mu^2}{m^2_B}}
-(\frac{13}{2} + \ln{r_1}) \ln{\frac{\mu^2_f}{m^2_B}}
+\frac{7}{16} \ln^2{(x_1 x_2)}+ \frac{1}{8} \ln^2{x_1} \non
&&+ \frac{1}{4} \ln{x_1} \ln{x_2}
+ \left (- \frac{1}{4}+ 2 \ln{r_1} + \frac{7}{8} \ln{\eta} \right ) \ln{x_1}
+ \left (- \frac{3}{2} + \frac{7}{8} \ln{\eta} \right) \ln{x_2} \non
&&+ \frac{15}{4} \ln{\eta} - \frac{7}{16} \ln^{2}{\eta}
+ \frac{3}{2} \ln^2{r_1} - \ln{r_1}
+ \frac{101 \pi^2}{48} + \frac{219}{16} \Biggr ].
\label{eq:ffnlot2}
\eeq
The factor $F^{(1)}_{\rm T3}(x_i,t,q^2)$ in Eqs.~(\ref{eq:ffnlop},\ref{eq:ffnlo0})
denotes the NLO twist-3 contribution as defined  in Eq.~(\ref{eq:nlohk2}):
\beq
F^{(1)}_{\rm T3}(x_i,t,q^2)&=&\frac{\alpha_s(\mu_f) C_F}{4 \pi}
\Biggl [\frac{21}{4} \ln{\frac{\mu^2}{m^2_B}}
- \frac{1}{2}(6 + \ln{r_1}) \ln{\frac{\mu^2_f}{m^2_B}}
+ \frac{7}{16} \ln^2{x_1} - \frac{3}{8} \ln^2{x_2} \non
&& \hspace{-1cm}+ \frac{9}{8} \ln{x_1} \ln{x_2}
+ \left (- \frac{29}{8}+ \ln{r_1} + \frac{15}{8} \ln{\eta} \right ) \ln{x_1}
+ \left (- \frac{25}{16} + \ln{r_2} + \frac{9}{8} \ln{\eta} \right) \ln{x_2} \non
&&\hspace{-1cm}+ \frac{1}{2} \ln{r_1} - \frac{1}{4} \ln^{2}{r_1} + \ln{r_2}
- \frac{9}{8} \ln{\eta} - \frac{1}{8} \ln^{2}{\eta} + \frac{37 \pi^2}{32}
+ \frac{91}{32} \Biggr ].
\eeq

\begin{figure*}
\centering
\vspace{-1cm}
\includegraphics[width=0.45\textwidth]{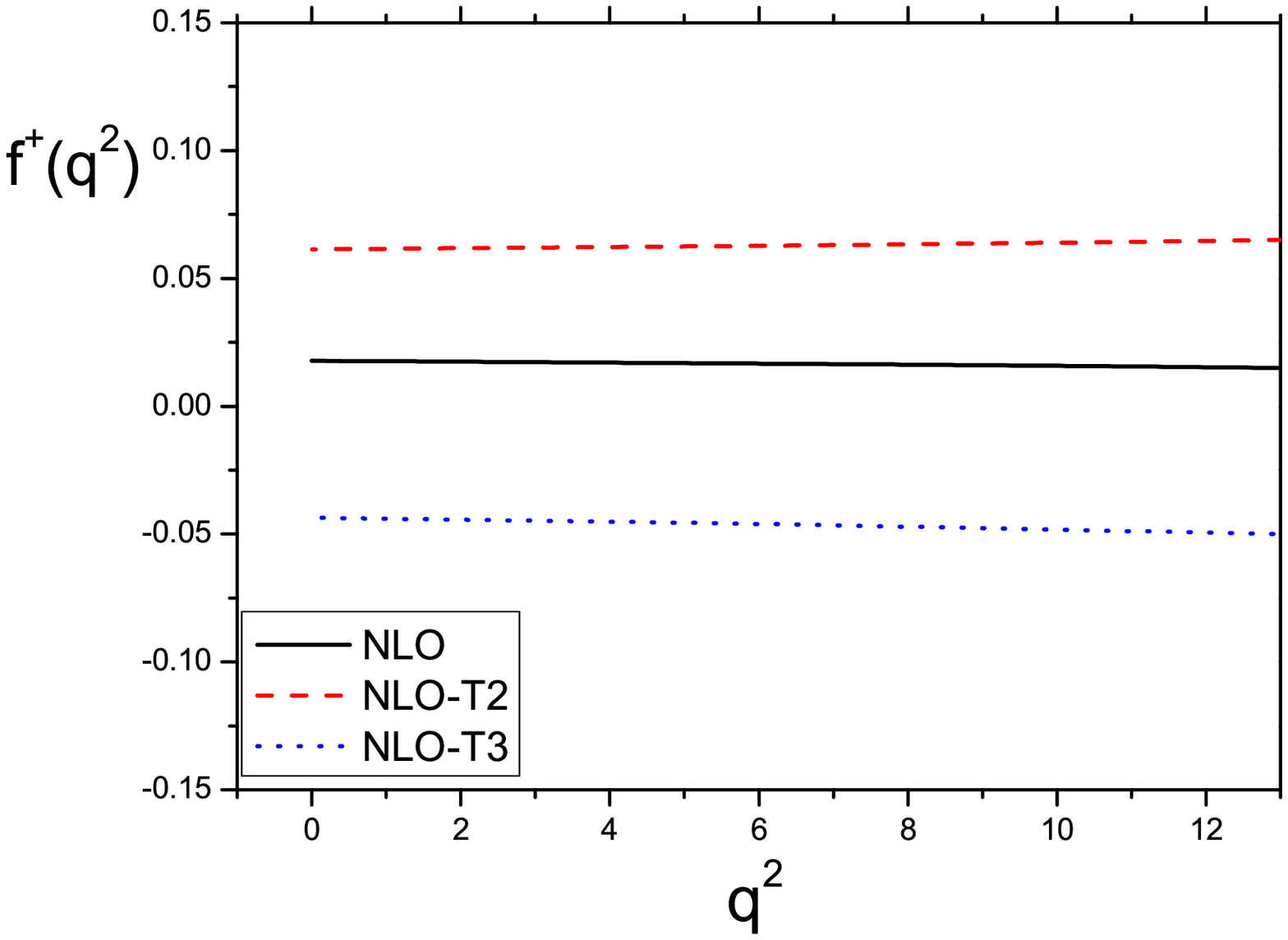}
\includegraphics[width=0.45\textwidth]{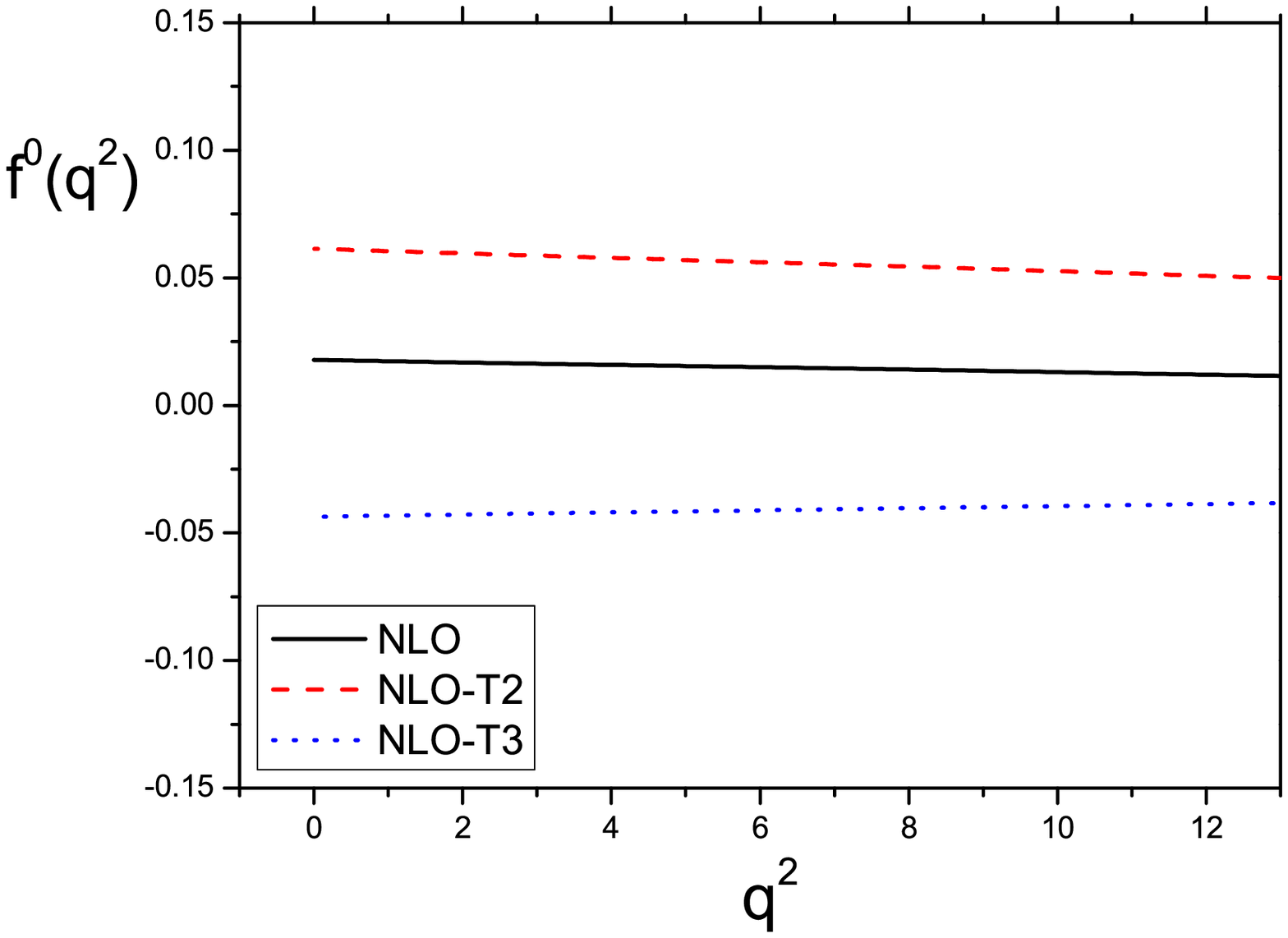}
\vspace{0cm}
\caption{The NLO twist-2 contribution (dashed-curve), the NLO twist-3 contribution
(dots-curve), and the total NLO contribution (the solid curve)
for $0\leq q^2 \leq 12$ GeV$^2$, and setting
$\mu_f=t$ and $\mu=t_s(\mu_f)$ as given in Eqs.~(\ref{eq:maxti},\ref{eq:scaleren}).}
\label{fig:fig7}
\end{figure*}
\begin{figure*}
\centering
\vspace{0cm}
\includegraphics[width=0.45\textwidth]{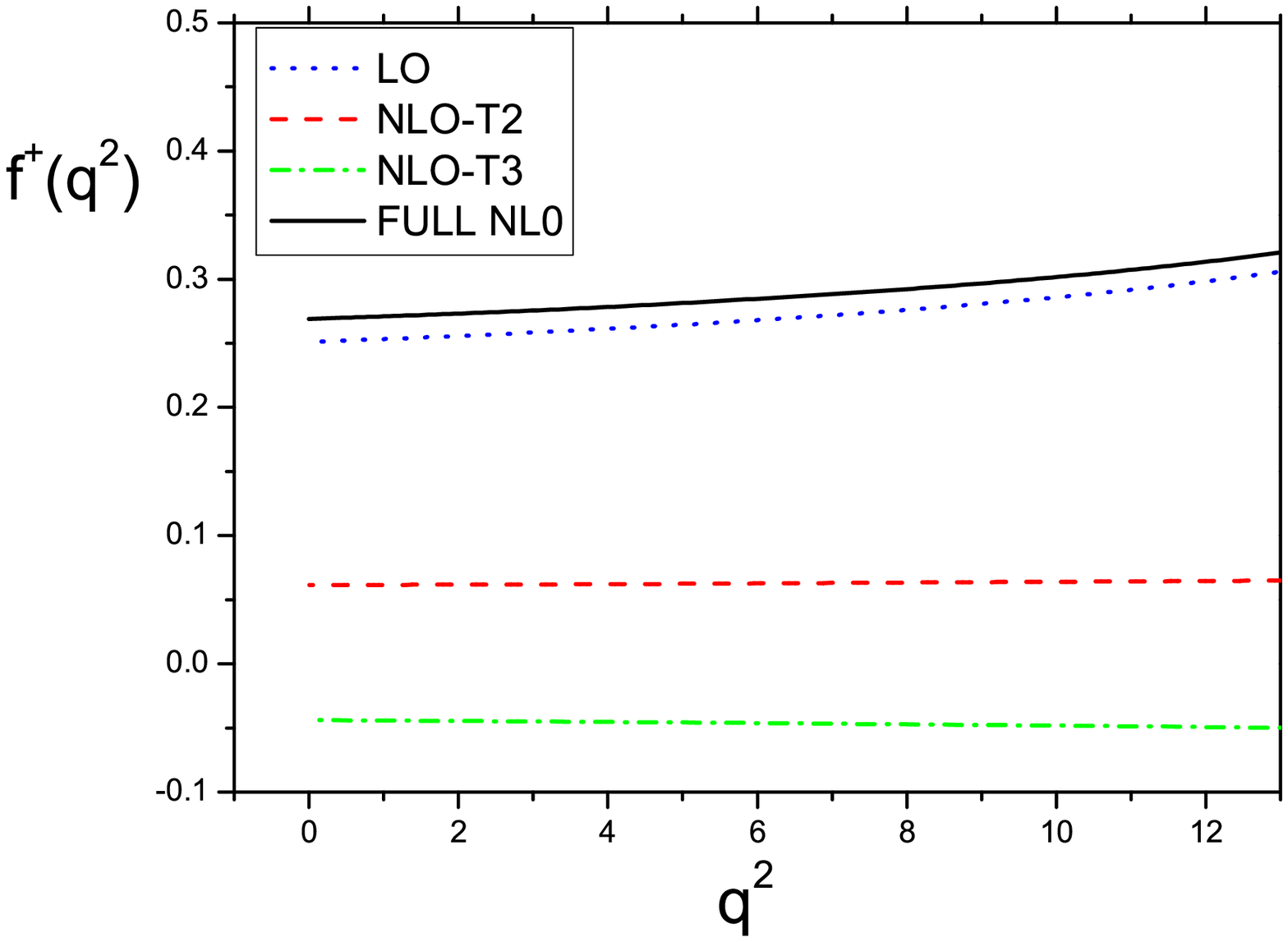}
\includegraphics[width=0.45\textwidth]{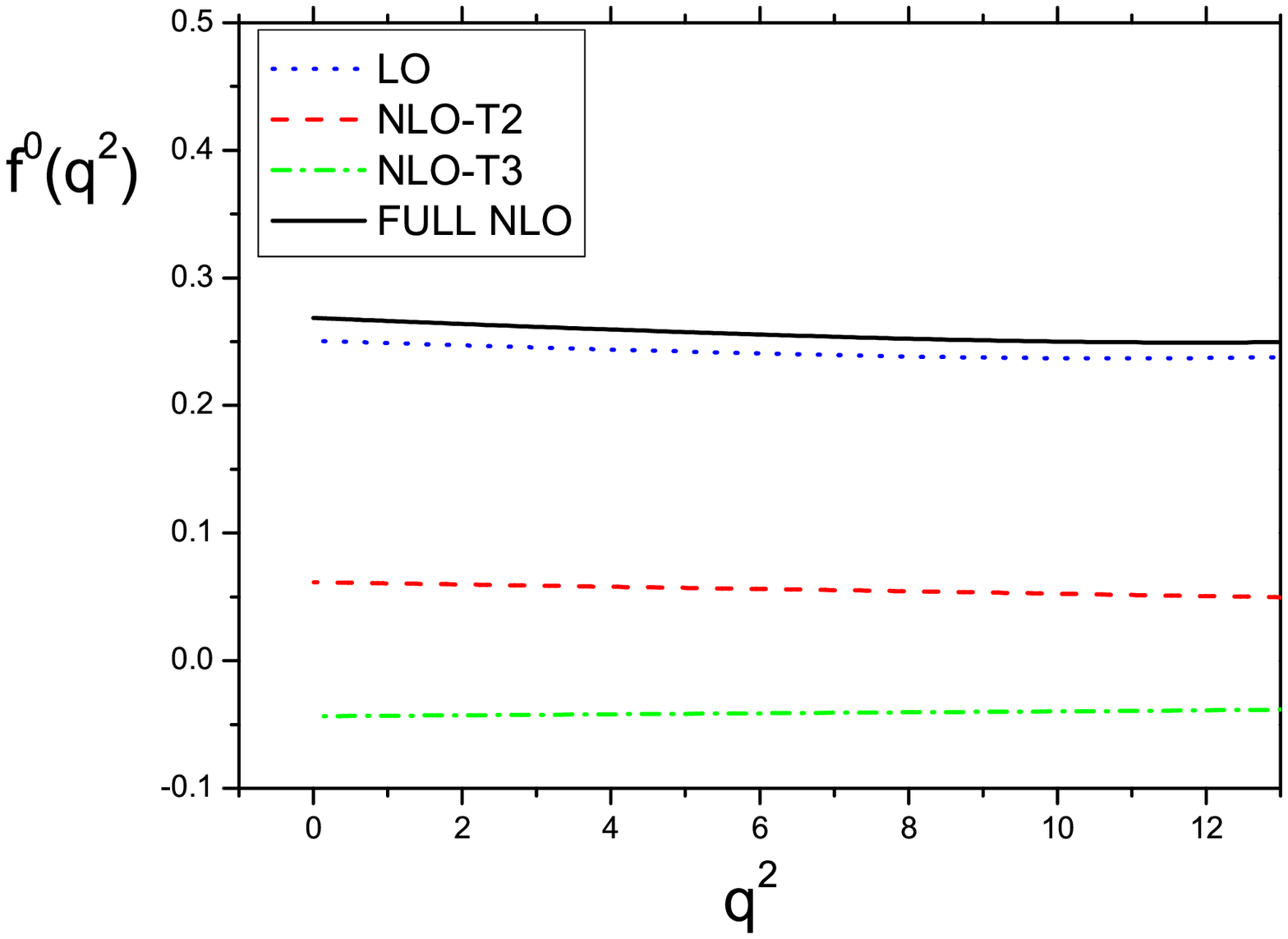}
\vspace{0cm}
\caption{The pQCD predictions for the form factor $f^+(q^2)$ and $f^0(q^2)$,
assuming $\omega_B=0.4$ and $c=0.3$, and setting $\mu_f=t$ and $\mu=t_s(\mu_f)$.
The left (right) diagram shows the $q^2$-dependence of the form factor,
with the inclusion of the full LO contribution (dots-curve), the NLO twist-2 contribution
only (dashed-curve), the NLO twist-3 one only (the dot-dashed curve), and finally
the total contribution at NLO level ( the solid curve).}
\label{fig:fig8}
\end{figure*}

\begin{table}[thb]
\begin{center}
\caption{ The pQCD predictions for the values of $f^+(q^2)$ and $f^0(q^2)$ for
$\omega_B=0.40$ and $c=0.3$, and assuming $q^2=(0,1,3,5,7,10,12)$ GeV$^2$.
The label LO, NLO-T2, NLO-T3, and NLO means the
full LO contribution, the NLO twist-2 part only, the NLO twist-3 part only, and
total contribution at NLO level: full LO plus both NLO twist-2 and twist-3 ones, respectively. }
\label{tab:table1}
\vspace{0.1cm}
\begin{tabular}{l | cccc ccc }
\hline \hline
$f^+(q^2)    $&$ 0   $&$ 1    $&$ 3    $&$5     $&$7     $&$10    $&$12   $  \\ \hline
LO    &$ 0.251 $&$ 0.254$&$ 0.257$&$0.266 $&$0.275 $&$0.285 $&$0.301$  \\
NLO-T2&$ 0.061 $&$ 0.061$&$ 0.062$&$0.063 $&$0.063 $&$0.064 $&$0.064$  \\
NLO-T3&$-0.043 $&$-0.044$&$-0.044$&$-0.045$&$-0.046$&$-0.047$&$-0.048$  \\
NLO   &$ 0.269 $&$ 0.271$&$ 0.275$&$0.284 $&$0.294 $&$0.302 $&$0.317$  \\ \hline\hline
$f^0(q^2)    $&$ 0     $&$ 1    $&$ 3    $&$ 5    $&$7     $&$10    $&$12   $  \\  \hline
LO     &$ 0.251 $&$ 0.248$&$ 0.246$&$0.243 $&$0.239 $&$0.237 $&$0.236$  \\
NLO-T2 &$ 0.061 $&$ 0.060$&$ 0.059$&$0.057 $&$0.055 $&$0.052 $&$0.051$  \\
NLO-T3 &$-0.043 $&$-0.043$&$-0.042$&$-0.042$&$-0.041$&$-0.040$&$-0.039$  \\
NLO    &$ 0.269 $&$ 0.265$&$ 0.263$&$0.258 $&$0.253 $&$0.249 $&$0.248$  \\ \hline\hline
\end{tabular}
\end{center}
\end{table}
\begin{table}[thb]
\begin{center}
\caption{The pQCD predictions for various contributions $f^+_{i}(q^2)$ for $q^2=(0,5,10)$
and their ratios $R_i(q^2)=f^+_i(q^2)/f^+_{LO}(q^2)$.}
\label{tab:table2}
\vspace{0.1cm}
\begin{tabular}{l| c  c|cc|cc}
\hline \hline
Source &$f_{i}(0)$& $R_i(0)$  &$f_{i}(5)$ & $R_i(5)$  &$f_{i}(10)$ & $R_i(10)$  \\ \hline
LO     &$0.251$   &$100 \%$ &$0.266$   &$100 \%$ &$0.285$   &$100 \%$     \\ \hline
LO-T2  &$0.086$   &$34.3\%$ &$0.084$   &$31.6\%$ &$0.082$   &$28.8\%$     \\
NLO-T2 &$0.061$   &$24.3\%$ &$0.063$   &$23.7\%$ &$0.064$   &$22.5\%$     \\ \hline
LO-T3  &$0.165$   &$65.7\%$ &$0.182$   &$68.4\%$ &$0.203$   &$71.2\%$     \\
NLO-T3 &$-0.044$  &$-17.1\%$&$-0.045$  &$-16.9\%$&$-0.047$  &$-16.5\%$    \\ \hline
NLO &$0.269$   &$107.2\%$&$0.284$   &$106.8\%$&$0.302$   &$106.0\%$    \\ \hline \hline
\end{tabular}
\end{center}
\end{table}

The $q^2$-dependence of the form factor $f^+(q^2)$ and $f^0(q^2)$ in the $\kt$
factorization up to NLO are shown in Fig.~\ref{fig:fig7} and Fig.~\ref{fig:fig8}.
In order to show and compare directly the relative strength of the contributions
from different sources, we also list the pQCD predictions for the
values of $f^+(q^2)$ and $f^0(q^2)$ in Table ~\ref{tab:table1},
assuming $\omega_B=0.40$, $c=0.3$ and $q^2=(0,1,3,5,7,10,12)$ GeV$^2$ respectively.
In Table ~\ref{tab:table1}, the label ``LO", ``NLO-T2", ``NLO-T3", and ``NLO"
mean the full LO contribution ( LO twist-2 plus LO twist-3), the NLO twist-2 part
only, the NLO twist-3 part only, and the total contribution at the NLO level
( full LO contribution plus both NLO twist-2 and NLO twist-3 ones), respectively.
In Table ~\ref{tab:table2}, for the cases of $f^+(q^2)$ with $q^2=(0,5,10)$ GeV$^2$,
we show the pQCD predictions for various contributions
to $f^+(q^2)$ from different sources: the LO twist-2, LO twist-3,
NLO twist-2, NLO twist-3, and finally the total contribution at the NLO level.
We also define the ratios $R_i=f_i^+(q^2)/f^+_{\rm LO}(q^2)$ to measure
the relative percentage of different contributions with respect to
the full LO contribution.

From the curves in  Fig.~\ref{fig:fig7} and Fig.~\ref{fig:fig8} and the
numerical results in Table ~\ref{tab:table1} and ~\ref{tab:table2},
one can have the following observations:
\begin{enumerate}
\item[(i)]
The NLO corrections at twist-2 and twist-3 are both under control, about
$20\%$ of the full LO contributions.
The reason is that the end-point region of $x_1$ is strongly suppressed and
the large double logarithm $\ln^2{x_1}$ in $H^{(1)}$ don't bring the dominant
contribution in the NLO corrections at both twists.

\item[(ii)]
From Fig.~\ref{fig:fig7} and Table I and II, one can see that
the NLO twist-2 and NLO twist-3 contributions are similar
in size but have an opposite sign, which leads to a strong cancelation
between NLO twist-2 and NLO twist-3 contributions and consequently
results in a small total NLO contribution, as illustrated explicitly in Fig.~\ref{fig:fig8}.
For the case of $f^+(0)$, for example, the LO twist-2 contribution is roughly
half of the LO twist-3
part: $34\%$ and $66\%$ of the full LO contribution respectively,
while the NLO twist-2 contribution can provide a $\sim 24\%$
enhancement to the LO prediction, but the NLO twist-3 part
can provide a $\sim 17.5\%$ decrease for the LO one.
The total NLO contribution results in, consequently,
a $7\%$ enhancement to the LO $f^+(0)$ only.

\item[(iii)]
Since the pQCD calculation for the form factors is reliable only at low $q^2$
region, we therefore show the pQCD predictions for $f^+(q^2)$ and $f^0(q^2)$
in the region of $ 0 \leq q^2 \leq 12$ GeV$^2$ only. One can see from
Figs.~\ref{fig:fig7} and \ref{fig:fig8} that the pQCD predictions for the two
form factors have a weak $q^2$-dependence: a $24\%$ ($22\%$) increase for
the LO (NLO) prediction for $f^+(q^2)$, but a $8\%$ ($6\%$) decrease for
the LO (NLO) prediction for $f^0(q^2)$, for the variation of $q^2$ from $q^2=0$
to $q^2=12$ GeV$^2$.
\end{enumerate}

\begin{figure*}[t,b]
\centering
\includegraphics[width=0.45\textwidth]{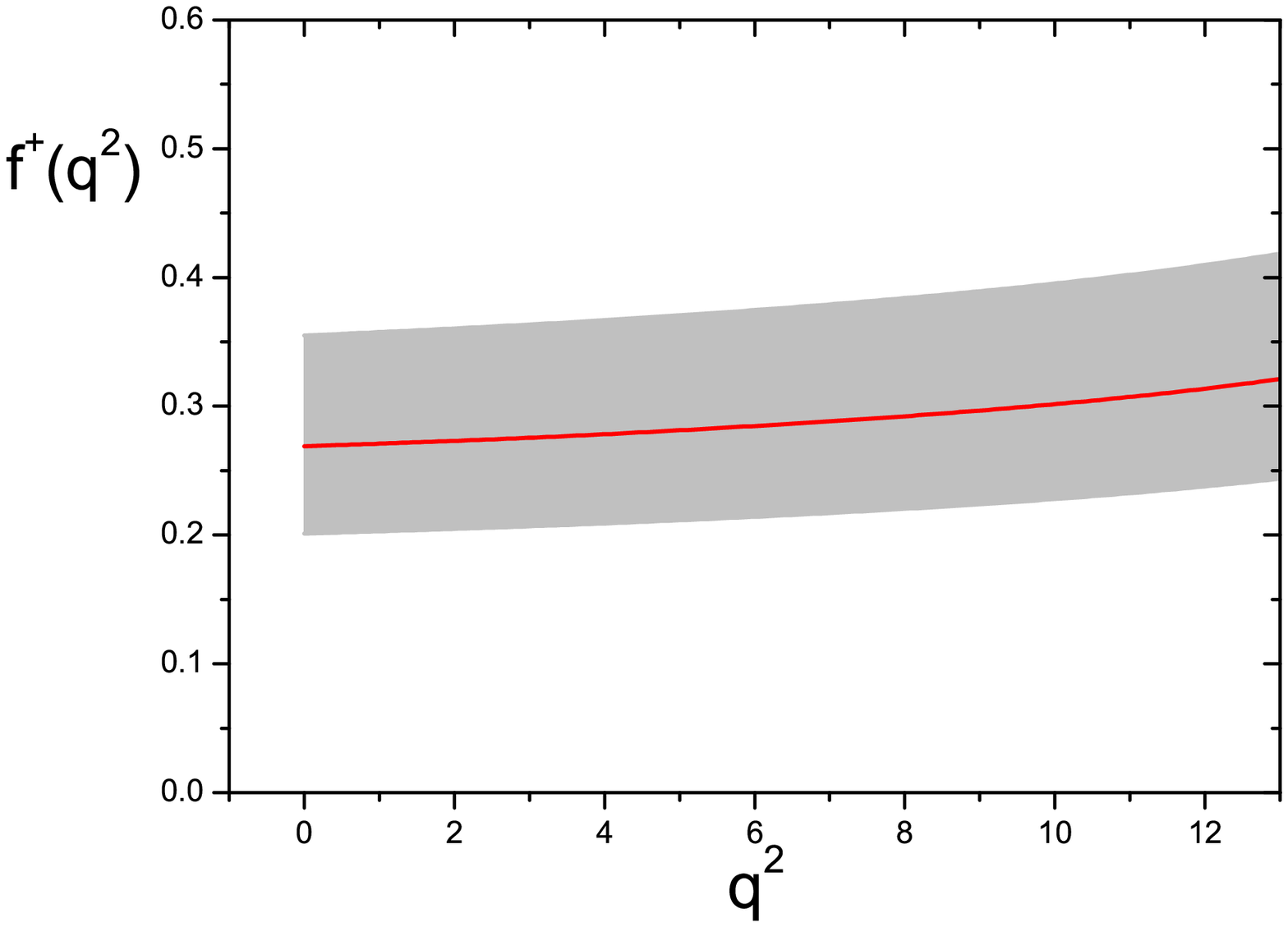}
\includegraphics[width=0.45\textwidth]{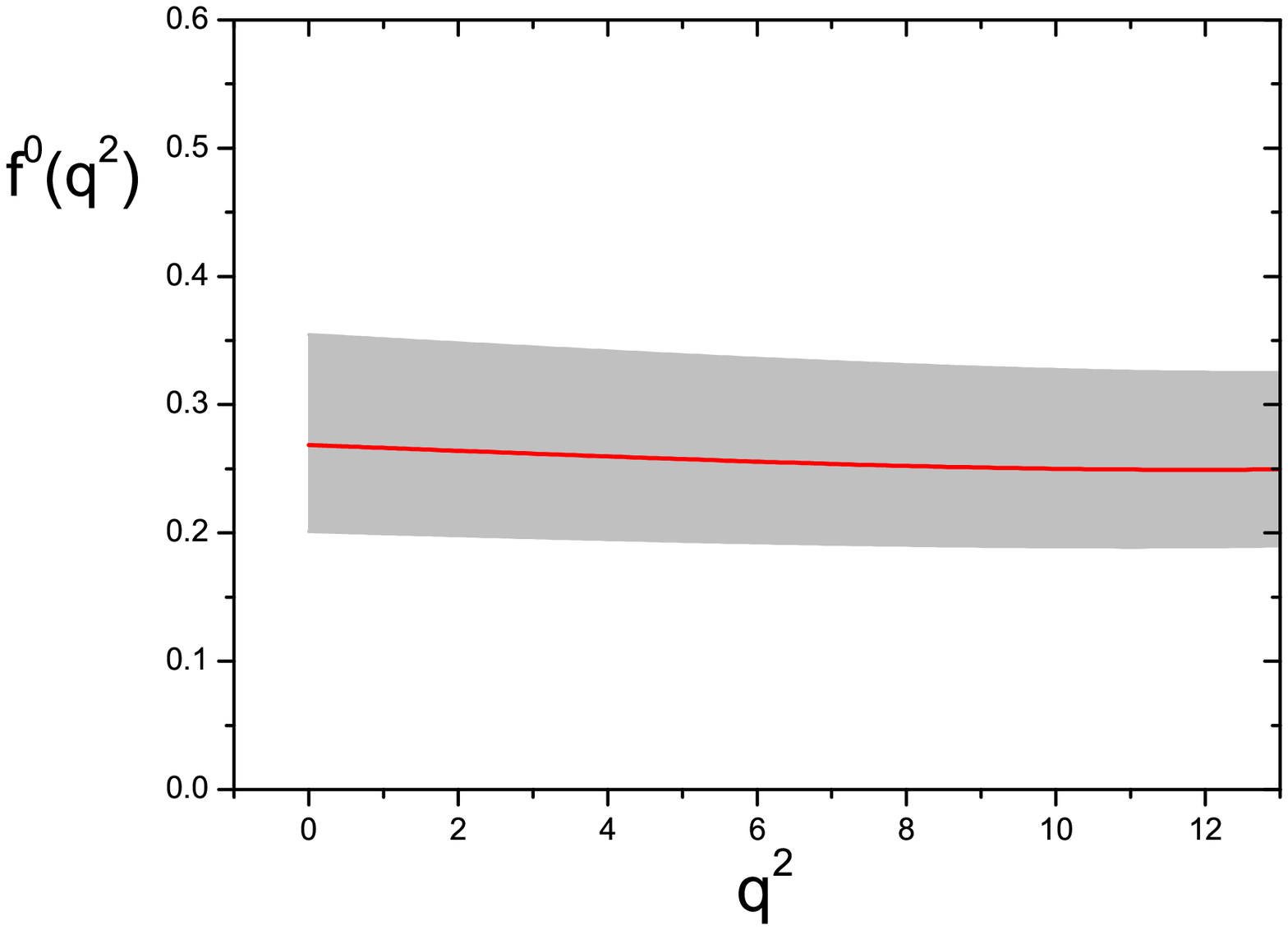}
\vspace{0cm}
\caption{Theoretical uncertainties of the $B \to \pi$ transition form
factor with the choice $\mu_f=t$ and $\mu=t_s(\mu_f)$ in the range of
$0\leq q^2 \leq 12$ GeV$^2$.}
\label{fig:fig9}
\end{figure*}

In our numerical calculations, the main theoretical errors
come from the uncertainties of the input parameters
$\omega_B=0.40\pm0.04$, $a_2=0.25\pm0.15$ and $m^{\pi}_0=1.4\pm 0.2$ GeV.
In Fig.~\ref{fig:fig9}, we show the central values and the theoretical uncertainties
of the NLO pQCD predictions for both form factors $f^+(q^2)$ and $f^0(q^2)$ of
$B \to \pi$ transition with the input hadron distribution amplitudes expressed
in Eqs.~(\ref{eq:dapion},\ref{eq:daB}), where the theoretical errors from different sources
are added in quadrature. For the case of $q^2=0$ ($\eta=1$), we find numerically that
\beq
f^+(0)&=&f^0(0)=0.269 ^{+0.042}_{-0.035}(w_B) ^{+0.028}_{-0.029}
(a_2^\pi) \pm 0.020 (m_0^{\pi})\non
& =& 0.269^{+0.054}_{-0.050}. \label{eq:ffat0}
\eeq
It is easy to see that the total theoretical error of the NLO pQCD
prediction for $f^{+,0}(0)$ is about $20\%$ of its central value, and keep
stable for the whole range of $0\leq q^2 \leq 12$ GeV$^2$, as
illustrated in Fig.~\ref{fig:fig9}.

In the previous numerical evaluations, we have used most popular choices for both
B meson~\cite{prl91-102001,prd69-034014,jhep0804-061,jhep0905-091,jhep1110-069}
and pion distribution amplitudes~\cite{plb87-359,prd22-2157,plb94-245,prl43-545}:
the B meson DA's  as shown in Eq.~(\ref{eq:daB}) with the relation of
$\phi_B=\phi_B^+=\phi_B^-$ and the non-asymptotic pion DA's $\phi_\pi^A$
and $\phi_\pi^{P,T}$ as given in Eq.~(\ref{eq:dapion}). We denote this set of
choices for B and pion DA's as the Case-A: $\phi_B(x,b) \oplus \phi_\pi^{A,P,T}$.

In Ref.~\cite{prd85-074004}, besides the Case-A, the authors also
considered other cases by using another form of B-meson DA inspired by the QCD sum rule \cite{prd55-272}:
$\phi_B^{\rm II}=(\phi_B^+ +\phi_B^-)/2$ with different $\phi_B^+$ and $\phi_B^-$:
\beq
\phi_B^{(+)} &=& \frac{f_B}{2 \sqrt{6}} x \left (\frac{m_B}{\omega_{B}}\right )^2
                \cdot \exp\left [-\frac{x m_B}{\omega_B} - \frac{1}{2} (\omega_B b)^2
                \right ], \non
\phi_B^{(-)} &=& \frac{f_B}{2 \sqrt{6}} \left (\frac{m_B}{\omega_{B}}\right )
                \cdot \exp\left [-\frac{x m_B}{\omega_B} - \frac{1}{2} (\omega_B b)^2 \right ],
\label{eq:daB1}
\eeq
as well as the asymptotic pion DA's $\phi_\pi^{asy}$:
\beq
\phi_{\pi}^{A}(x) =\frac{3 f_{\pi}}{ \sqrt{6}} x (1-x), \quad
\phi_{\pi}^{P}(x) =\frac{f_{\pi}}{2 \sqrt{6}}, \quad
\phi_{\pi}^{T}(x) =\frac{f_{\pi}}{2 \sqrt{6}} (1-2 x).
\label{eq:dapion1}
\eeq

Following Ref.~\cite{prd85-074004}, we here will also make the numerical
calculations for other three possible ways of choices of B and pion meson DA's:
\beq
{\rm Case-B:}~~~ \phi_B \oplus \phi_\pi^{asy};\quad
{\rm Case-C:}~~~ \phi_B^{\rm II} \oplus \phi_\pi;\quad
{\rm Case-D:}~~~ \phi_B^{\rm II} \oplus \phi_\pi^{asy}.
\eeq
We will compare the numerical results obtained for the different cases.

\begin{figure*}[t,b]
\centering
\includegraphics[width=0.45\textwidth]{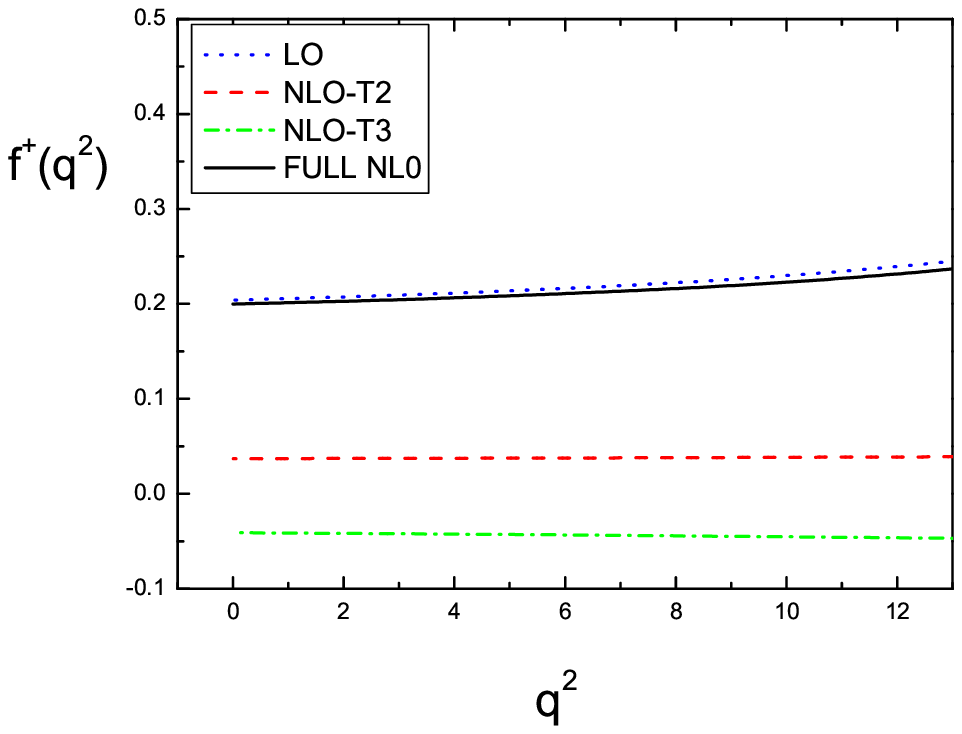}
\includegraphics[width=0.45\textwidth]{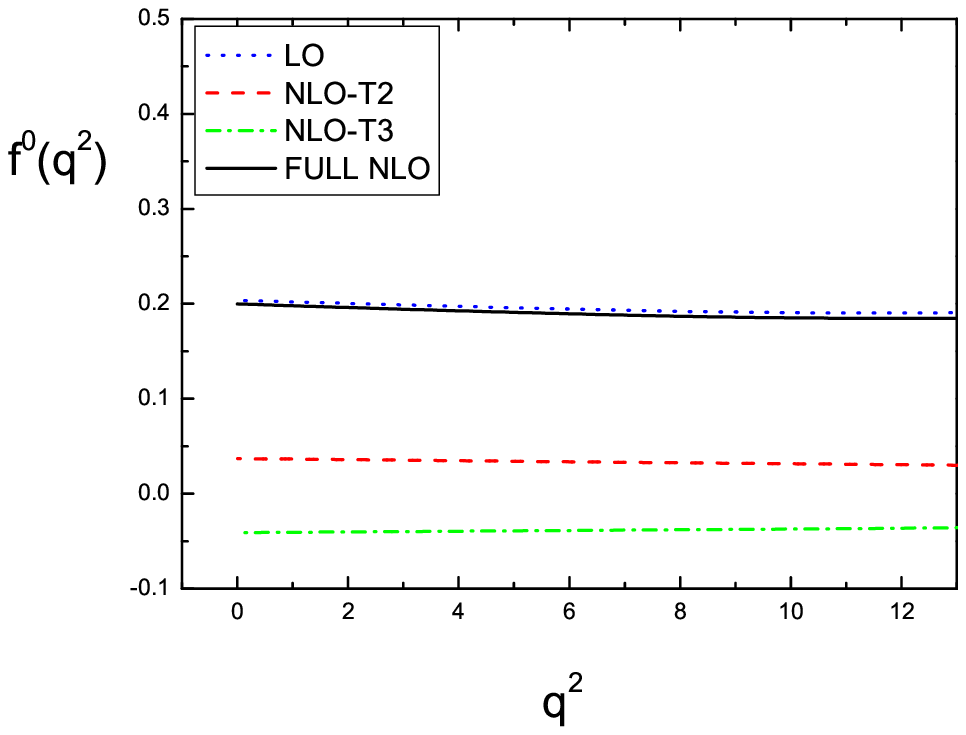}
\vspace{0cm}
\caption{The pQCD predictions for the form factors $f^+(q^2)$ and $f^0(q^2)$
for the Case-B:  the B meson DA in Eq.~(\ref{eq:daB}) and the asymptotic pion DA's
in Eq.~(\ref{eq:dapion1}) are used in the numerical calculation. }
\label{fig:fig10}
\end{figure*}

In Fig.~\ref{fig:fig10}, firstly, we show the $q^2$ dependence of the form factors
for the Case-B: i.e. the pQCD predictions for $f^+(q^2)$ and $f^0(q^2)$ obtained
by using $\phi_B$ and $\phi_\pi^{asy}$ as given in Eq.~(\ref{eq:daB})and
Eq.~(\ref{eq:dapion1}), respectively. By this way, we can check  the impact of the
higher conformal-spin partial waves in pion DA's, which partially arose from the
nonzero pion mass correction.
From the curves in Fig.~\ref{fig:fig10}, one can see that both form factors $f^+(q^2)$
and $f^0(q^2)$ are reduced by about $20\%$ in the whole range of $ 0< q^2 < 12$ Gev$^2$
when the additional Gegenbauer terms in pion DA's are not included.
In Table ~\ref{tab:table3}, furthermore, we list the pQCD predictions for various
contributions to $f^+(q^2)$ for $q^2=(0,5,10)$ GeV$^2$ and their
ratios $R_i(q^2)=f^+_i(q^2)/f^+_{LO}(q^2)$
for the Case-B.
When compared with the numerical results for the Case-A as listed in
Table ~\ref{tab:table2}, we find that the NLO Twist-3 contribution
in Case-B plays a more important
role than that for the Case-A. For Case-B, the NLO Twist-2 and NLO
Twist-3 contribution largely
canceled each other, the full NLO form factor  then become a little smaller than the LO one.

\begin{table}[thb]
\begin{center}
\caption{The pQCD predictions for various contributions to $f^+_{i}(q^2)$ for $q^2=(0,5,10)$
and their ratios $R_i(q^2)=f^+_i(q^2)/f^+_{LO}(q^2)$ for the Case-B.}
\label{tab:table3}
\vspace{0.1cm}
\begin{tabular}{l| c  c|cc|cc}
\hline \hline
Source &$f_{i}(0)$& $R_i(0)$  &$f_{i}(5)$ & $R_i(5)$  &$f_{i}(10)$ & $R_i(10)$  \\ \hline
LO     &$0.204$   &$100  \%$  &$0.215$    &$100 \%$   &$0.230$   &$100  \%$     \\ \hline
LO-T2  &$0.065$   &$31.9 \%$  &$0.063$    &$29.3\%$   &$0.061$   &$26.5 \%$     \\
NLO-T2 &$0.037$   &$18.1 \%$  &$0.038$    &$17.7\%$   &$0.038$   &$16.5 \%$     \\ \hline
LO-T3  &$0.139$   &$68.1 \%$  &$0.152$    &$70.7\%$   &$0.168$   &$73.0 \%$     \\
NLO-T3 &$-0.041$  &$-20.1\%$  &$-0.043$   &$-20.0\%$  &$-0.045$  &$-19.6\%$     \\ \hline
NLO    &$0.20$   &$98.0 \%$  &$0.210$    &$97.7\%$   &$0.221$   &$96.1 \%$     \\ \hline \hline
\end{tabular}
\end{center}
\end{table}

\begin{figure*}[t,b]
\centering
\includegraphics[width=0.45\textwidth]{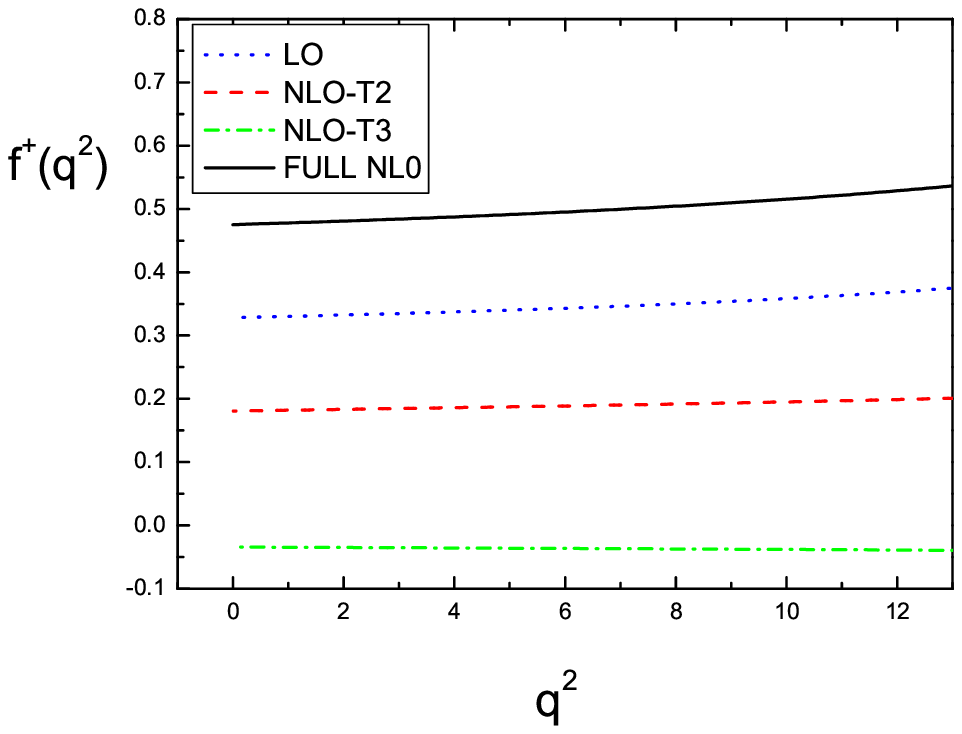}
\includegraphics[width=0.45\textwidth]{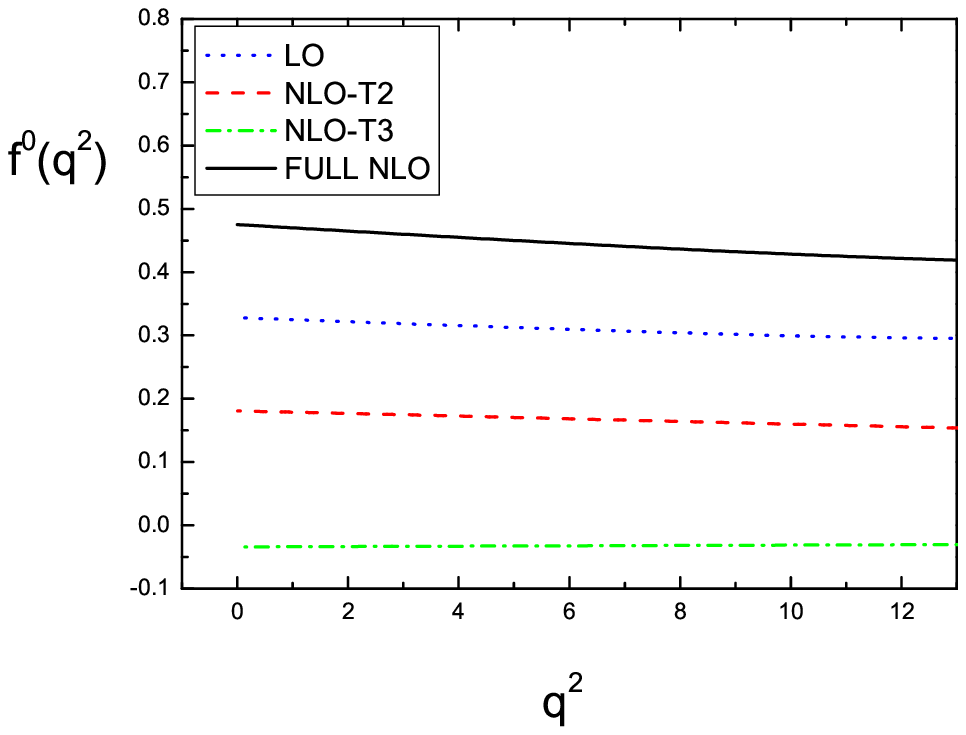}
\vspace{0cm}
\caption{The pQCD predictions for the form factors $f^+(q^2)$ and $f^0(q^2)$
for the Case-C: the B meson DA's as given in Eq.~(\ref{eq:daB1})  and
the non-asymptotic pion DA's in Eq.~(\ref{eq:dapion}) are used.}
\label{fig:fig11}
\end{figure*}

In Fig.~(\ref{fig:fig11}) and Table \ref{tab:table4}, we show the pQCD predictions for
the form factors $f^+(q^2)$ and $f^0(q^2)$, and for their ratios
$R_i(q^2)=f^+_i(q^2)/f^+_{LO}(q^2)$
for the Case-C: where the B meson DA's as given in Eq.~(\ref{eq:daB1})  and
the non-asymptotic pion DA's in Eq.~(\ref{eq:dapion}) are used.
The same input parameters as in Case-A are used here.
For the Case-C, we find that
\begin{itemize}
\item[(i)]
The LO contribution to form factors $f^+(0)$ and $f^0(0)$ is $0.328$ for Case-C, which is
much larger than $f_i(0)=0.251$ for Case-A, since the LO-T2 term for Case-C is $0.148$,
much larger than $0.086$ for Case-A.

\item[(ii)]
The net NLO contribution to form factors $f_i(0)$ is about $0.15$ for Case-C,
much larger than $0.017$ for Case-A, since the NLO-T2 term for Case-C is $0.181$
instead of the small $0.061$ for Case-A.

\item[(iii)]
When we take all four parts into account, we find a large NLO pQCD prediction:
$f_i(0)=0.475$ for Case-C, which is much larger
than $f_i(0)=0.269$ for case-A,  and also rather different from the popular values
obtained by using QCD sum rule.

\end{itemize}

\begin{table}[thb]
\begin{center}
\caption{The pQCD predictions for various contributions to $f^+_{i}(q^2)$ for $q^2=(0,5,10)$
GeV$^2$ and their ratios $R_i(q^2)=f^+_i(q^2)/f^+_{LO}(q^2)$ for the Case-C.}
\label{tab:table4}
\vspace{0.1cm}
\begin{tabular}{l| c  c|cc|cc}
\hline \hline
Source &$f_{i}(0)$& $R_i(0)$  &$f_{i}(5)$ & $R_i(5)$  &$f_{i}(10)$ & $R_i(10)$  \\ \hline
LO     &$0.328$   &$100  \%$  &$0.341$    &$100 \%$   &$0.358$   &$100  \%$     \\ \hline
LO-T2  &$0.148$   &$45.1 \%$  &$0.146$    &$42.8\%$   &$0.145$   &$40.5 \%$     \\
NLO-T2 &$0.181$   &$55.2 \%$  &$0.188$    &$55.1\%$   &$0.195$   &$54.5 \%$     \\ \hline
LO-T3  &$0.180$   &$54.9 \%$  &$0.195$    &$57.2\%$   &$0.213$   &$59.5 \%$     \\
NLO-T3 &$-0.034$  &$-10.4\%$  &$-0.036$   &$-10.6\%$  &$-0.038$  &$-10.6\%$     \\ \hline
NLO    &$0.475$   &$144.8\%$  &$0.493$    &$144.6\%$  &$0.515$   &$143.9\%$     \\ \hline \hline
\end{tabular}
\end{center}
\end{table}

\begin{figure*}[t,b]
\centering
\includegraphics[width=0.45\textwidth]{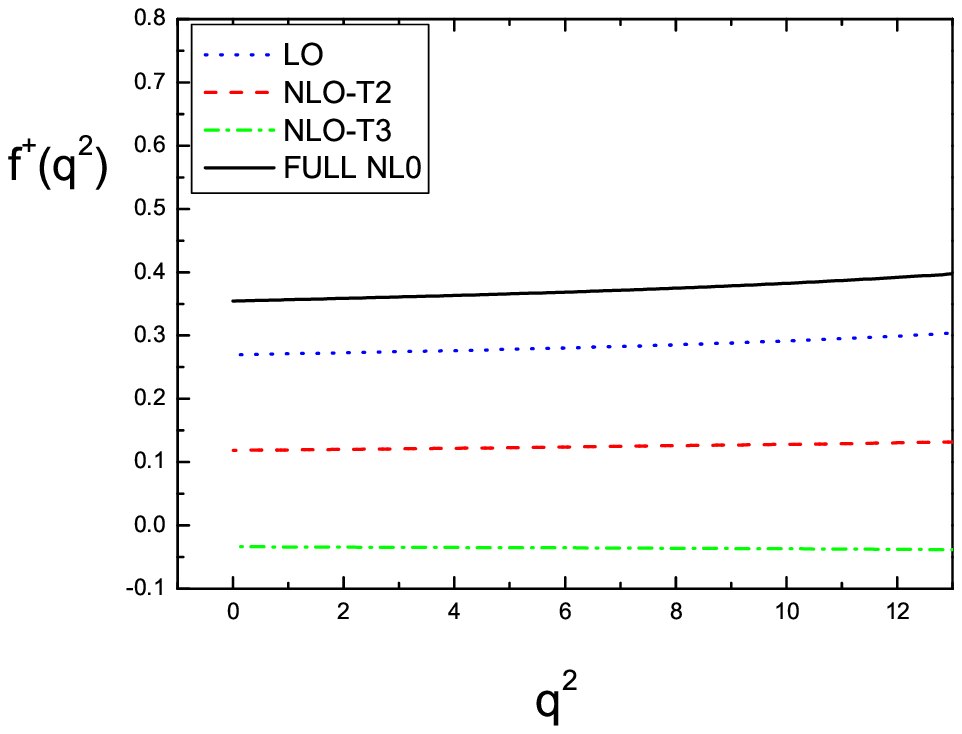}
\includegraphics[width=0.45\textwidth]{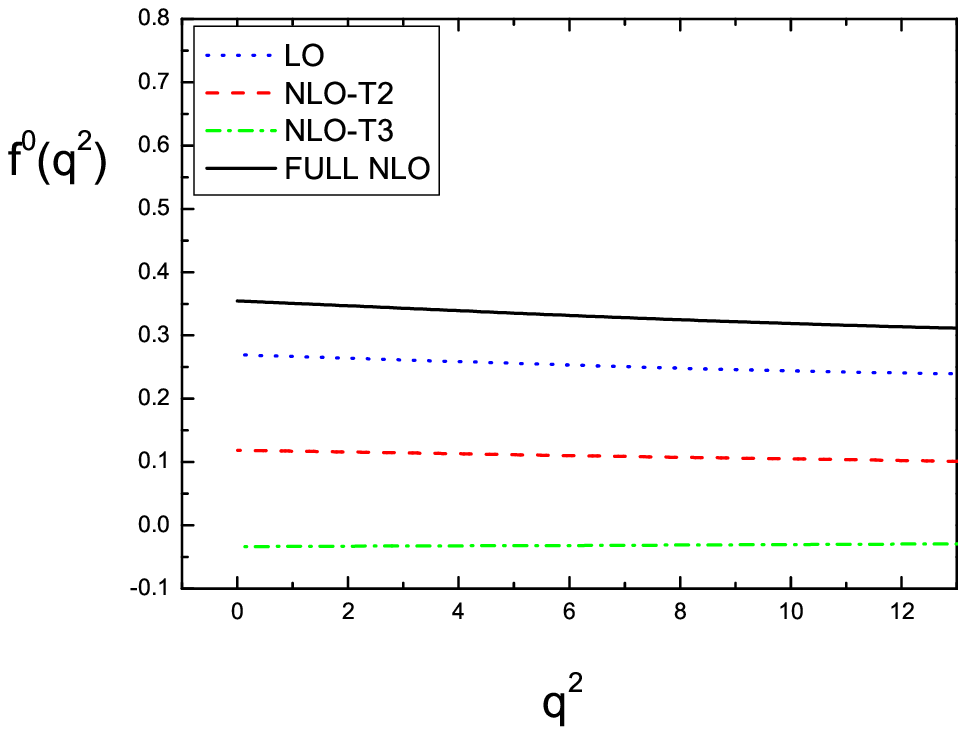}
\vspace{0cm}
\caption{The pQCD predictions for the form factors $f^+(q^2)$ and $f^0(q^2)$ for the Case-D.}
\label{fig:fig12}
\end{figure*}

In Fig.~(\ref{fig:fig12}) and Table \ref{tab:table5}, finally, we show the pQCD
predictions for the form factors $f^+(q^2)$ and $f^0(q^2)$, and for their ratios
$R_i(q^2)=f^+_i(q^2)/f^+_{LO}(q^2)$
for the Case-D: where the B meson DA's as given in Eq.~(\ref{eq:daB1})  and
the asymptotic pion DA's in Eq.~(\ref{eq:dapion1}) are used. The same input
parameters as in Case-A are used here.
For this case, both LO-T2 and NLO-T2 term become much larger than those for
Case-A, and lead to a large LO and NLO
pQCD predictions for $f_i(0)$. The NLO part here provides a $31\%$ enhancement
to the LO one.

\begin{table}[thb]
\begin{center}
\caption{The pQCD predictions for various contributions to $f^+_{i}(q^2)$ for
$q^2=(0,5,10)$ GeV$^2$ and to their ratios $R_i(q^2)=f^+_i(q^2)/f^+_{LO}(q^2)$
for the Case-D.}
\label{tab:table5}
\vspace{0.1cm}
\begin{tabular}{l| c  c|cc|cc}
\hline \hline
Source &$f_{i}(0)$& $R_i(0)$  &$f_{i}(5)$ & $R_i(5)$  &$f_{i}(10)$ & $R_i(10)$  \\ \hline
LO     &$0.270$   &$100  \%$  &$0.279$    &$100 \%$   &$0.290$   &$100  \%$     \\ \hline
LO-T2  &$0.115$   &$42.6 \%$  &$0.114$    &$40.9\%$   &$0.112$   &$38.6 \%$     \\
NLO-T2 &$0.118$   &$43.7 \%$  &$0.123$    &$44.1\%$   &$0.128$   &$44.1 \%$     \\ \hline
LO-T3  &$0.155$   &$57.4 \%$  &$0.165$    &$59.1\%$   &$0.178$   &$61.4 \%$     \\
NLO-T3 &$-0.034$  &$-12.6\%$  &$-0.035$   &$-12.5\%$  &$-0.037$  &$-12.8\%$     \\ \hline
NLO    &$0.354$   &$131.1\%$  &$0.367$    &$131.5\%$  &$0.381$   &$131.4\%$     \\ \hline \hline
\end{tabular}
\end{center}
\end{table}

\section{Summary}

In this paper, by employing the $\kt$ factorization theorem,
we calculated the NLO twist-3 contribution to the form factors $f^+(q^2)$
and $f^0(q^2)$ of the $B \to \pi$ transition.

The UV divergences are renormalized into the coupling constants, decay constant
and quark fields. Both the soft and collinear divergences in the NLO QCD quark
diagrams and in the NLO effective diagrams for meson wave functions are regulated
by the off-shell momentum $k^2_{\rm iT}$ of the light quark.
The heavy $b$ quark is protected on-shell to treat it as the standard effective
heavy quark field in the $\kt$ factorization theorem, and then the soft gluon
radiated by the $b$ quark can be regularized by the gluon mass $m_g$.
With the reasonable choice of $\xi_2^2=m^2_B$, only the NLO corrections of the
$B$ meson function develop an additional double logarithm $\ln^2{r_1}$, with
$r_1=\xi^2_1/m^2_B$. And then the resummation technique is implemented to
minimize the scheme dependence from the different choice of $\xi_1^2$.

The cancelation of the IR divergences between the QCD quark diagrams and
the effective diagrams for the meson wave functions at twist-3, in cooperation
with the cancelation at the leading twist, verifies the
validity of the $\kt$ factorization for the $B\to \pi l^-\bar{\nu}_l $ semileptonic
decays at the NLO level.
The large double logarithm $\ln^2{x_1}$ in the NLO hard kernel is resummed
to result in the Sudakov factor, while the single logarithms and
constant terms in the NLO hard kernel are all diminished by the choice
of the scale $\mu$ and $\mu_f$. We have demonstrated explicitly that the NLO
corrections are under control.

From our numerical evaluations, we generally find that the NLO pQCD predictions
for the form factors $f^+(q^2)$ and $f^0(q^2)$ for the Case-A agree well with
those obtained by using
the QCD some rule. Based on our calculations we find the following points:
\begin{itemize}
\item[(i)]
For Case-A, the full LO and NLO pQCD predictions are $f^{+,0}_{\rm LO}(0)=0.251$
and $f^{+,0}_{\rm NLO}(0)=0.269$,
which is consistent with those from QCD sum rule.

\item[(ii)]
There is a strong cancelation between the NLO twist-2 and NLO twist-3 contribution
to the form factors $f^{+,0}(q^2)$ of $B \to \pi$ transition.
For the case of $f^+(0)$, for example, the NLO twist-2 contribution
provides roughly $24\%$ enhancement to the full LO one, but the NLO twist-3 contribution
makes a $17.5\%$ decrease for the full LO result.
The total NLO contribution results in a $7\%$ enhancement to the LO pQCD prediction,
which is small and stable for the whole range of $0\leq q^2\leq 12$ GeV$^2$.

\item[(iii)]
For other three cases, i.e. using different choices of the B meson and pion DA's in our numerical evaluations,
the LO and NLO pQCD predictions will change accordingly.
Generally speaking, the pQCD predictions for Case-C and Case-D  are much larger  than those
obtained from the QCD sum rule.

\end{itemize}

\begin{acknowledgments}
The authors would like to thank H.N.~Li, Y.L.~Shen, W.F.~Wang and X.~Liu for valuable discussions.
This work is supported by the National Natural Science Foundation of China
under Grant No.11235005, 11228512 and 11375208;
and by the Project on Graduate Students¡¯ Education and Innovation of Jiangsu
Province under Grant No. CXZZ13-0391.

\end{acknowledgments}

\end{document}